\documentclass[a4paper,11pt]{article}
\pdfoutput=1
\usepackage{jcappub-alt}
\usepackage[utf8]{inputenc}
\usepackage[english]{babel}
\usepackage{physics}
\usepackage{bbold}
\usepackage{amssymb,amsmath,amsfonts}
\usepackage{csquotes}
\usepackage{fancyvrb}
\usepackage{tensor}
\usepackage{layout}
\usepackage[normalem]{ulem}

\usepackage{multirow}
\usepackage{verbatim}
\usepackage{comment}
\usepackage{scalefnt}
\usepackage[usenames,dvipsnames]{xcolor}

\usepackage{graphicx,color}

\usepackage{orcidlink}

\usepackage{subcaption}
\usepackage{soul}

\definecolor{urlblue}{rgb}{0,0,0.9}\definecolor{linkblue}{rgb}{0,0,.8}\definecolor{linkgreen}{rgb}{0,0.45,0}\definecolor{linkpurple}{rgb}{0.7,0.0,0.4}\definecolor{linkorange}{rgb}{0.7,0.1,0.0}\AtBeginDocument{\hypersetup{linkcolor=linkblue,citecolor=linkorange,urlcolor=urlblue}}\definecolor{urlblue}{rgb}{0,0,0.9}\definecolor{linkblue}{rgb}{0,0,.8}\definecolor{linkgreen}{rgb}{0,0.45,0}\definecolor{linkpurple}{rgb}{0.7,0.0,0.4}\definecolor{linkorange}{rgb}{0.7,0.1,0.0}

\newcommand{\nn}{\mathcal{N}}
\newcommand{\mpc}{$[{\rm Mpc}/h]^3$}

\author[a, f, g]{Adrian P. Schirra\orcidlink{0009-0004-9589-5579},}
\author[b, c, d, e]{Miguel Quartin\orcidlink{0000-0001-5853-6164},}
\author[a]{Luca Amendola\orcidlink{0000-0002-0835-233X}}

\affiliation[a]{Institute of Theoretical Physics, Heidelberg University, Philosophenweg 16, 69120 Heidelberg, Germany}

\affiliation[b]{Instituto de Física, Universidade Federal do Rio de Janeiro, 21941-972, Rio de Janeiro, RJ, Brazil}

\affiliation[c]{Observatório do Valongo, Universidade Federal do Rio de Janeiro, 20080-090, Rio de Janeiro, RJ, Brazil}

\affiliation[d]{PPGCosmo, Universidade Federal do Espírito Santo, 29075-910, Vitória, ES, Brazil}

\affiliation[e]{Centro Brasileiro de Pesquisas Físicas, 22290-180, Rio de Janeiro, RJ, Brazil}

\affiliation[f]{Institute of Nuclear and Particle Physics, TUD Dresden University of Technology, 01062 Dresden, Germany}

\affiliation[g]{Deutsches Zentrum für Astrophysik, Postplatz 1, 02826 Görlitz, Germany}

\emailAdd{adrian.schirra@dzastro.de}
\emailAdd{mquartin@cbpf.br}
\emailAdd{l.amendola@thphys.uni-heidelberg.de}

\title{
A model-independent measurement of the expansion and growth rates from BOSS using the FreePower method}

\abstract{
    In this work we provide a data analysis of the BOSS galaxy clustering data with the recently proposed FreePower method, which  adopts as parameters the power spectrum band-powers, the expansion rate, and the growth rate instead of  specific cosmological parametrizations. It relies on the Alcock-Paczyński effect and redshift-space distortions, and makes use of one-loop perturbation theory for biased tracers. In this way, we obtain for the first time constraints on the linear growth rate, on the Hubble parameter, as well as on the dimensionless distance $H_0 D_A$ and various bias functions, that are independent of a model for the power spectrum shape and thus of both the early and late-time cosmological modelling.  Using weakly-informative priors, requiring basically that $\sigma_8 \in [0.67, 1.07]$ at 95\% CI, we find at $z_{\rm eff}=0.38$, $f=0.67^{+0.20}_{-0.19}$, $H/H_0=1.033^{+0.13}_{-0.081}$, $H_0 D_A = 0.264^{+0.026}_{-0.039}$ and at $z_{\rm eff}=0.61$, $f=0.82^{+0.25}_{-0.20}$, $H/H_0=1.085^{+0.16}_{-0.067}$, $H_0 D_A = 0.390^{+0.036}_{-0.046}$. We find lower $H/H_0$ results than expected from Planck 2018 $\Lambda$CDM results at a confidence level of 1.7$\sigma$ ($2.1\sigma$) for low-$z$ (high-$z$).
    These results form a proof-of-principle of the FreePower method.
    We also get constraints on the bias parameters which are in agreement with constraints from previous BOSS analyses, which serves as a cross-check of our pipeline.
}

\begin{document}

\maketitle

\section{Introduction}

Ongoing  surveys will soon provide extensive datasets probing the distribution of galaxies on very large volumes. Important ground surveys (DESI~\cite{DESI:2024uvr,DESI:2024lzq}, 4MOST~\cite{2019Msngr.175....3D}, J-PAS~\cite{2021A&A...653A..31B}, Vera C. Rubin Observatory LSST~\cite{2009arXiv0912.0201L}) and space surveys (Euclid \cite{2018LRR....21....2A,2011arXiv1110.3193L}, Nancy Grace Roman Space Telescope~\cite{2021MNRAS.507.1514E,2021arXiv211103081R}) of the large-scale structure (LSS) of the Universe are  already underway or scheduled to commence shortly. The corresponding increase in precision of these surveys calls for an accurate theoretical modeling.

As it has become clear in the last decade or so, a treatment using linear cosmological perturbation theory is no longer sufficient to optimize the science return  \cite{Cabass:2022avo}. An understanding of the non-linear formation of structures and the astrophysical uncertainties is necessary to use the observational data in the best  possible way~\cite{2022arXiv220906854G}. Perturbation theory provides the means to investigate LSS observables in the weakly non-linear regime. A comparison between the model predictions from non-linear perturbation theory and survey data can therefore lead to new insights. The theoretical higher-order power spectrum that we employ in this work has been derived within the context of the Effective Field Theory of the Large Scale Structure (EFTofLSS)~\cite{Baumann:2010tm,Carrasco:2012cv,Ivanov:2019pdj,DAmico:2019fhj} (see also~\cite{Cabass:2022wjy,Philcox:2021kcw} for the bispectrum and \cite{Chudaykin:2020ghx,Ivanov:2020ril,Xu:2021rwg,Lague:2021frh} for non-standard cosmologies). EFTofLSS makes it possible to absorb the short-distance physics, which is not known in detail \cite{2022arXiv220906854G}, into a set of parameters which can then be fitted to data. For an alternative approach based on Kinetic Field Theory, see e.g.~\cite{Lilow:2018ejs}.

The EFTofLSS studies fluctuations in the density of biased tracers, such as galaxies. Its natural  cutoff lies at a scale where the gravitational evolution becomes highly non-linear and the impact of astrophysical processes becomes very high. Below the cutoff, EFT provides a connection between the initial conditions after inflation and the observables in the late Universe \cite{Cabass:2022avo}.

Complementing the EFTofLSS treatment, the impact of small-scale physics on the long-wavelength fluctuations can be modeled by ultraviolet (UV) counterterms \cite{Ivanov:2019pdj}. They are also necessary because the short-scale physics is not modeled by perturbation theory. Moreover, infrared (IR) resummation \cite{2015JCAP...02..013S} has to be applied to account for the fact that the shape of the baryon acoustic oscillation (BAO) peak is very sensitive to long-wavelength modes (bulk flows), which cannot be treated perturbatively either. Furthermore, to establish a connection between the theoretical modeling of the dark matter density contrast and the observables, it becomes necessary to consider the galaxy bias~\cite{Desjacques:2016bnm}. This provides a connection between the dark matter density field and the galaxy number density field, and redshift space distortions (RSD) \cite{Perko:2016puo}, which account for the fact that the galaxies have peculiar velocities due to clustering dynamics.

Most cosmological analyses of real data assume a (standard or non-standard) cosmological model in the theoretical description (see in particular  \cite{Ivanov:2019pdj,Philcox:2021kcw,DAmico:2019fhj} for the BOSS data). However, the results one  gets are bound to depend on the underlying cosmological model, and cannot be employed to test different scenarios. It is therefore desirable to pursue an alternative route in order to remain as model-independent as possible \cite{Samushia:2010ki,Amendola:2019lvy,Boschetti:2020fxr,Amendola:2022vte}. In this way, one reduces the chances to miss new physics or to introduce biases in the parameter estimation.

An approach to remain model independent, which was recently proposed in~\cite{Amendola:2019lvy,Amendola:2022vte,Amendola:2023awr}, consists in dividing the linear matter power spectrum $P(k)$ into several $k$-bins whose values are free to vary. From these bins, and a set of bias and counterterm parameters, the non-linear galaxy power spectrum and bispectrum are derived, allowing comparisons with the galaxy survey data. Since the Alcock-Paczyński (AP) effect~\cite{1979Natur.281..358A,DAmico:2019fhj} changes the multipole structure of the RSD (assuming an incorrect cosmology results in changes to the multipole amplitudes), and this structure is fundamental and independent of the early-universe or dark energy models, the correct cosmology can be inferred as that which recovers this fundamental multipole structure. In this case, it is not necessary to make assumptions about the shape of $P(k)$. At the same time, the expansion rate $H(z)$ and the growth rate $f(z)$ are left free to vary with redshift. The non-linear correction to one loop in the power spectrum are evaluated assuming very general kernels, derived under the assumption of a homogeneous and isotropic background and the equivalence principle of general relativity, which implies that one can remove a pure-gradient metric perturbation by going to the free-falling frame of comoving observers~\cite{Desjacques:2016bnm}. All the non-linear parameters are left free to vary in each redshift. In this way, the results are independent of the details of both  early-time models (that determine the initial power spectrum shape) and  late-time models (that determine the background and perturbations growth). In Ref.~\cite{Amendola:2023awr} this framework was denoted as the FreePower method.

The growth rate $f$ should also be binned in $k$-space since in several modified gravity models it depends on $k$  but for this first real-data analysis this proves too demanding and for simplicity we assume the growth to be scale independent, in line with most similar analyses. In Ref.~\cite{Amendola:2022vte} we have shown that if $f$ depends on $k$ then one needs at least three multipoles to apply our methodology and recover all the parameters. One can show that taking $f$ to be $k$-independent, two multipoles are sufficient, which simplifies our analysis.

An interesting aspect of the FreePower method is that we can measure the dimensionless expansion rate $E$ and dimensionless angular diameter distance $L_A$, defined respectively by
\begin{equation}
    E\equiv H(z)/H_0\,, \qquad L_A\equiv H_0 D_A(z)\,,
\end{equation}
independently of a model for the background or for the power spectrum shape. This is possible because both the power spectrum and bispectrum are distorted by the 
AP effect in a way that depends on $E,L_A$ (see Appendix~\ref{app:one-loop} for more information). Another remarkable aspect of FreePower is that although the linear bias $b_1$ and the growth rate  $f$ are degenerate with the linear spectrum shape $P(k)$, this degeneracy is broken at the non-linear level. The same occurs for $E$ and $L_A$. Using this we have shown, in particular, how FreePower can, without assuming any specific cosmological model, estimate the spatial curvature~\cite{Amendola:2024gkz}, or be combined with supernova and gravitational wave distances to test for the presence of modified gravity or cosmic opacity~\cite{Matos:2023jkn}.

To obtain constraints on the various cosmological and bias parameters using observational data, we  make use of Markov Chain Monte Carlo (MCMC) methods. A fast numerical evaluation of the non-linear galaxy power spectrum at the next to leading order (i.e.~the one-loop order) is necessary for this procedure. This can be achieved with the FFTLog method~\cite{2018JCAP...04..030S}, where the linear $P(k)$ is expanded as a superposition of power-law functions. The loop calculations can then be done analytically.  An implementation of this method is provided by the code \texttt{PyBird}~\cite{DAmico:2020kxu}.

The main aim of this work is to produce the first real-data analysis with the FreePower framework using data from the Baryon Oscillation Spectroscopic Survey (BOSS). We do not strive for the best possible precision in this first real-data application of FreePower. Instead, we focus on demonstrating that the method works and is able to produce competitive constraints for $E(z)$ and $f(z)$. Since each likelihood evaluation is expensive and, as discussed below, we will have 17 free parameters, we had difficulties with the large computational cost of the MCMC codes  when using a small, 256-core computing cluster. We thus adopted two main simplifications with respect to~\cite{Amendola:2023awr} in order to reduce the total computational time and get reliable results in a reasonable amount of time. First, we do not include the bispectrum, which, if included, would result in increased precision~\cite{Amendola:2023awr}. Second, we adopt the analytical  kernels for an Einstein-deSitter (EdS) model. EdS kernels should in any case be a very good approximation for cosmologies that do not depart too much from $\Lambda$CDM; they have been adopted also in a previous analysis of BOSS data that we will compare to \cite{Ivanov:2019pdj}.

With the above simplifications, we obtained competitive constraints on cosmological and non-cosmological parameters. In particular, we are able to measure the dimensionless expansion rate and the growth function at both BOSS redshifts. This study also paves the way for model-independent analyses of future observational data of the LSS.

\section{The FreePower model-independent approach}

\subsection{Theoretical model\label{sec:tm}} The theoretical model for the galaxy density field power spectrum is given as \cite{Ivanov:2019pdj,DAmico:2020kxu}
\begin{equation}
    P_{\rm gg}(k,\mu,z)=P_{\rm 11}(k,\mu,z)+P_{\rm 22}(k,\mu,z)+P_{\rm 13}(k,\mu,z)+P_{\rm ctr}(k,\mu,z)+P_{\rm shot}.\label{eq:ourmodel}
\end{equation}
The summands are the tree-level galaxy power spectrum, the two one-loop corrections, the counterterms, and the shot noise.

In this application of the FreePower method, the theoretical power spectrum (\ref{eq:ourmodel}) depends on a number of free parameters at each redshift (see Table~\ref{tab:priors1}): the background functions $E(z),L_A(z)$ that enter the Alcock-Paczyński effect; the linear growth rate $f(z)$; the linear and non-linear bias parameters $b_1, b_2,b_{\mathcal{G}_2}$; the counterterms $c_0,c_2$; and the shot noise $s$ (see App.~\ref{app:one-loop} for the definition of these parameters and more details regarding the individual terms).

\begin{figure}
    \centering
    \includegraphics[width=.7\columnwidth]{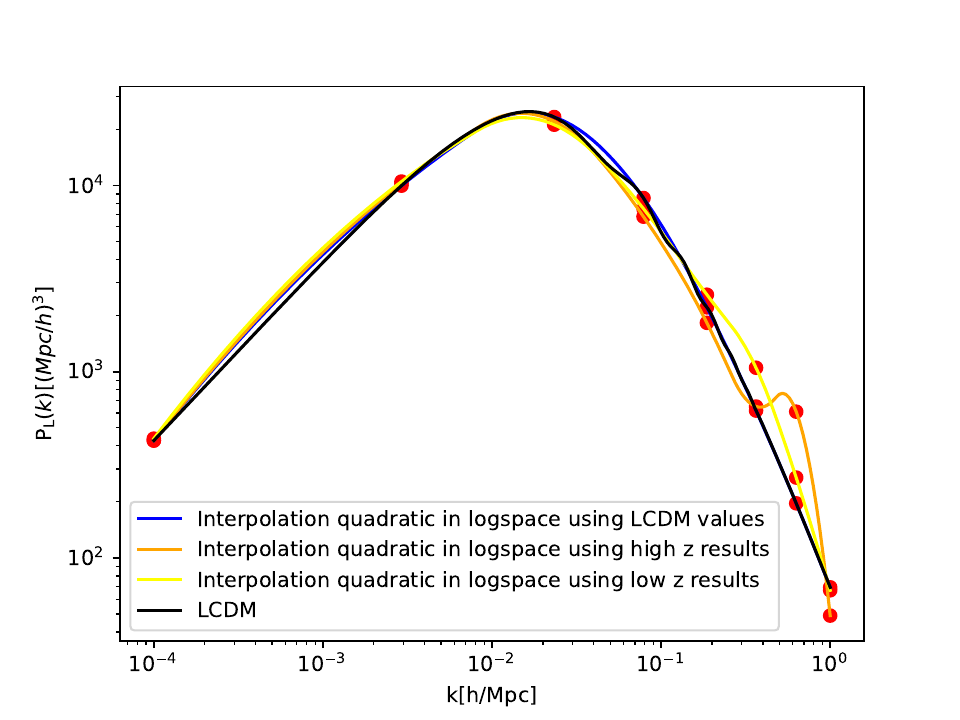}
    \caption{
    The linear $P(k)$ in the FreePower method is obtained as an interpolation between the 8 best-fit variable parameters $P_i$ (red points). In black, we depict a test $\Lambda$CDM $P_L(k)$, and in blue the corresponding reconstructed $P_L(k)$ through interpolation. In orange (yellow) we show instead the reconstructed $P_L(k)$ with our best-fit $P_i$ for the high (low) redshift bins.
    }
    \label{fig:P_L_interpolation}
\end{figure}

Importantly, the model~\eqref{eq:ourmodel} uses the linear matter power spectrum which, in standard analyses, is  calculated within specific cosmological models.
In contrast, in the model-independent approach presented here, the linear matter power spectrum is parametrized in $k$-bins and the waveband values are varied in the MCMC. We select the values of $P_L$ at the central $k$-values
\begin{equation}
    \{0.0001, 0.003, 0.023, 0.079, 0.187, 0.364, 0.630, 1.0\}\,\, h/{\rm Mpc}.\label{eq:wb}
\end{equation}
The linear matter power spectrum must be provided in this range to calculate the one-loop correction integrals in \texttt{PyBird} \cite{DAmico:2020kxu}, following the standard procedure of EFTofLSS of using scales beyond $k_{\rm max}$ in the computation of the loop integrals~\cite{Chudaykin:2020aoj,DAmico:2022osl}. In this way, we introduce eight new parameters to vary in the MCMC analysis, but there is no underlying assumption about the cosmological model when the linear matter power spectrum is calculated anymore. Previous tests indicate that neither the exact spacing nor the number of the $k$-bins has a strong influence on the results on the other parameters~\cite{masterarbeit3}. The linear matter power spectrum is constructed with a spline interpolation between these varied parameters. To decide the best interpolation scheme, we computed the 8 $k$-bins assuming $\Lambda$CDM and found out that using a spline which is second order in log space  yielded the best results -- see Appendix~\ref{app:interpolation} for more details.  So this is the interpolation scheme we adopt in this work. We stress that the interpolation does not contain the BAO wiggles nor any other $\Lambda$CDM feature.
Figure~\ref{fig:P_L_interpolation} illustrates the spline fits for the $\Lambda$CDM test-set, as well as for the best fit $P_i$ we found below for both low and high-$z$ bins.

Because of that, the linear matter power spectrum is quite smooth. Therefore, the coupling between the large-scale IR modes and the mildly non-linear modes is rather weak~\cite{Cabass:2022avo} and the effect of IR resummation is in general small. In this approach we do not disentangle the growth factor $G(z)$ from the linear power spectrum, so we in fact measure $P_L(k,z)=G(z)^2P_L(k)$ at each redshift bin.

In Ref.~\cite{Amendola:2023awr} we also included an overall smoothing factor that takes into account the Finger-of-God effect and the redshift uncertainty; we found however that the effect of this smoothing was subdominant and here, therefore, for simplicity, we do not include it.

\subsection{Likelihood}

For every set of parameters, the multipoles of the galaxy power spectrum according to the model can now be calculated. To compare the model  to the data, we employ the likelihood~\cite{Beutler:2021eqq}
\begin{equation}
    \mathcal{L}\propto\exp\Bigg[-\frac{1}{2} \sum_{\ell  = {0, 2}}\sum_{\ell' = {0, 2}}\Big(\mathbf{P}_{\rm dat, \ell}-(WM\mathbf{P}_{\rm theo})_\ell \Big)^T C^{-1}_{\ell \ell'}\Big(\mathbf{P}_{\rm dat, \ell'}-(WM\mathbf{P}_{\rm theo})_{\ell'} \Big)\Bigg]\,.
\end{equation}
The full BOSS dataset contains five different multipoles: $\ell=0,1,2,3,4$ in the $k$-range $0-0.4 \,h/{\rm Mpc}$. It has 40 $k$-bins and a total vector of length $200$ (40 for each of the five multipoles). $W$ is a 200$\times$2000 survey window function matrix and $M$ is the 1200$\times$2000 wide-angle transformation matrix which transforms 3 multipoles into the 5 final multipoles~\cite{Beutler:2021eqq}.
$\mathbf{P}_{\rm theo}$ is a vector that contains the even multipoles ($\ell = 0, 2, 4$) of the model in $k$-bins of $0.001 \,h/{\rm Mpc}$. For our analysis, we choose $k_{\rm max}=0.3 \,h/{\rm Mpc}$ as our default limit. Large volume simulations have shown that one-loop is still reasonable up to this scale for BOSS data~\cite{Ivanov:2019pdj}. We have checked that using a more conservative $k_{\rm max} = 0.25 \,h$/Mpc, as in~\cite{Ivanov:2019pdj}, results in negligible shifts of the posterior peaks and only a slight widening of the contours, so our present results are not very sensitive to either choice.\footnote{We remark that any analysis should be limited to $k<0.3 \,h/{\rm Mpc}$ due to the Nyquist-frequency in the measured power spectra \cite{Beutler:2021eqq}.}
We also neglect $\ell = 1, \,3$ and $4$: we drop the odd multipoles and the hexadecapole both for simplicity, and in order to compare more closely to the results of~\cite{Ivanov:2019pdj}. The odd multipoles are often neglected in the literature, and the signal-to-noise ratio in the hexadecapole is very low compared to monopole and quadrupole~\cite{Simon:2022lde}.\footnote{hexadecapole functionality was also absent from \texttt{PyBird} until very recently, already in the late stages of our analysis.} So our $\mathbf{P}_{\rm dat}$ contains the data for the power spectrum multipoles $\ell=0,2$ in the $k$-range $0-0.3 \,h/{\rm Mpc}$ in bins of $0.01 \,h/{\rm Mpc}$.

The covariance matrix $C$ is estimated from mock catalogs of BOSS DR12 \cite{2016MNRAS.456.4156K,Beutler:2021eqq}. The Hartlap correction factor~\cite{2007A&A...464..399H} which could affect the covariance matrix at the $\approx 1\%$ level, is neglected here following ~\cite{Ivanov:2019pdj}. Furthermore, one could perform a rescaling of parameter errors to account for a bias caused by the mock based covariance estimate~\cite{Beutler:2021eqq,2013PhRvD..88f3537D,2014MNRAS.439.2531P}. However, this correction is at subpercent level too and is neglected here for simplicity.

\subsection{Data sets from BOSS}

\begin{table}
    \setlength\tabcolsep{6pt}
    \centering
    \begin{tabular}{lcrr}
        \hline\hline
         parameter name & symbol & prior low-$z$ sample & high-$z$ sample\\
         \hline
         linear growth rate                 & $f$         & ${\cal U}[0,1.5]$   & ${\cal U}[0,1.5]$   \\
         dimensionless expansion rate       & $E$         & ${\cal U}[0,2]$   & ${\cal U}[0,2]$  \\
         dim.less angular diameter distance & $L_A$       & ${\cal U}[0.12,0.38]$ & ${\cal U}[0.16,0.48]$\\
         linear bias                        & $b_1$       & $\nn(1.67,0.72)$  & $\nn(2.28,1.04)$   \\
         ${\cal G}_2$ bias     & $b_{\mathcal{G}_2}$      & $\nn(0.11,1.36)$  & $\nn(0.6,2.8)$        \\
         quadratic bias                     & $b_2$       & $\nn(-2.3,2.8)$   & $\nn(-2.7,5.2)$       \\
         dimensionless shot noise           & $s$         & $\nn(0,0.5)$      & $\nn(0,0.5)$       \\
         counterterm (monopole) & $c_0/[{\rm Mpc}/h]^2$   & $\nn(-18.5,95.7)$ & $\nn(31.5,116)$        \\
         counterterm (quadrupole) & $c_2/[{\rm Mpc}/h]^2$ & $\nn(14.7,72)$    & $\nn(-43.3,192)$       \\
         \hline
                                            & $P_1/$\mpc  & $\nn(429,42.9)$   & $\nn(429,42.9)$    \\
                                            & $P_2/$\mpc  & $\nn(9870,987)$   & $\nn(9870,987)$    \\
                                            & $P_3/$\mpc  & $\nn(23000,4000)$ & $\nn(23000,4000)$  \\
         linear power spectrum $k-$bins     & $P_4/$\mpc  & $\nn(8350,2000)$  & $\nn(8350,2000)$   \\
                                            & $P_5/$\mpc  & $\nn(2320,800)$   & $\nn(2320,800)$    \\
                                            & $P_6/$\mpc  & $\nn(911,500)$    & $\nn(911,500)$      \\
                                            & $P_7/$\mpc  & $\nn(375,500)$     & $\nn(375,500)$     \\
                                            & $P_8/$\mpc  & $\nn(100,200)$     & $\nn(100,200)$     \\
         \hline\hline
    \end{tabular}
    \caption{Free parameters used here in our  model-independent approach, and their prior ranges in the MCMC analysis. We follow in part~\cite{Simon:2022lde} in deciding the priors, adding a cut to the $P_i$ priors not to allow negative values.
    }
    \label{tab:priors1}
\end{table}

We consider here data from the Baryon Oscillation Spectroscopic Survey (BOSS), which was part of the Sloan Digital Sky Survey (SDSS) III~\cite{2011AJ....142...72E, 2013AJ....145...10D}. It contains data on anisotropic galaxy clustering in Fourier space.\footnote{\url{https://fbeutler.github.io/hub/deconv_paper.html} (last access: 22.03.2025).}
The survey includes $1 198 006$ luminous galaxies over $10 252$ ${\rm deg}^2$ \cite{Beutler:2021eqq}. The area is divided into the North Galactic Cap (NGC) and the South Galactic Cap (SGC). The data can be split into two redshift bins defined by $0.2 < z < 0.5$ and $0.5 < z < 0.75$. The values of the effective redshift are $z_{\rm eff} = 0.38$ and $0.61$~\cite{Beutler:2021eqq}. We often refer to the samples simply as low-$z$ and high-$z$ survey, respectively.

The effective redshift can be used because the redshift bin is much smaller than the scale of variation of the growth factor $G(z)$ and the non-cosmological parameters. Therefore, the binning corresponds to evaluating all quantities at the same effective redshift \cite{DAmico:2019fhj}.

\subsection{Settings for our analysis}

Due to the time constraints on our available computing system, in this study only the (larger) NGC data set at both redshift bins is employed.

We compare the theoretical predictions of the one-loop power spectrum model to monopole and quadrupole of the data vector. We only calculate the $2\times 400$ values of monopole and quadrupole of the model vector $\mathbf{P}_{\rm theo}$ and only use the $2\times 40$ values of monopole and quadrupole of $WM\mathbf{P}_{\rm theo}$ to compare them to the data vector $\mathbf{P}_{\rm dat}$ in the likelihood. As discussed above, the odd multipoles and the hexadecapole in $\mathbf{P}_{\rm dat}$ are not used in the present analysis. Due to its low signal-to-noise, we set the model prediction for the hexadecapole to zero. A theoretical prediction for the hexadecapole is required to calculate the window function contributions, since the window function couples the different multipoles~\cite{Beutler:2021eqq}. Since only consider monopole and quadrupole in the likelihood, the effect of this choice should be very small (see appendix \ref{app:hexadecapole_matrix_mixing}). For the data comparison, the $k$-range is restricted to $0.01 \,h/{\rm Mpc} <k<0.3 \,h/{\rm Mpc}$.

The bias parameters $b_1$, $b_2$, and $b_{\mathcal{G}_2}$ are left free to vary, together with the counterterm parameters $c_0$ and $c_2$.
The bias parameter $b_{\Gamma_3}$ is set to zero, following ISZ20. An alternative would be to set it to non-zero using coevolution relations~\cite{Philcox:2021kcw,2022PhRvD.105f3512I}, since in principle it influences $P(k)$ for high $k$, or leave it free to be marginalized over. However, it is strongly degenerate with the parameter $b_{\mathcal{G}_2}$, and thus its effect can be largely absorbed by the $b_{\mathcal{G}_2}$ term. In fact, it was found in~\cite{Wadekar:2020hax} that, in the traditional full shape approach, only $b_{\mathcal{G}_2}$ was affected when setting $b_{\Gamma_3}$ to zero or not. We leave a possible cross-check of this result in the context of the FreePower method for future work.
The parameter $c_4$ need not be considered, since we do not use the hexadecapole here.

The priors of the varied parameters are presented in Table~\ref{tab:priors1}. The priors assume no correlation amongst any of our parameters. We also assume each redshift bin is completely independent, and no correlation exists among the two parameter sets.
The priors on the eight parameters describing the linear matter power spectrum, as well as the bias parameters and counterterms were originally intended to be very broad and non-informative.  However, our tests showed that very broad priors in our large multidimensional space lead to ``prior-volume effects'', introducing  spurious results after marginalization (see e.g.~\cite{DESI:2024jis,DESI:2024hhd}).
This issue, combined with the large computational time to run many different chains with different prior choices, prompted us to use priors which are not completely uninformative. For $P_3$--$P_8$ we set them by after trial and error, so that the likelihood had enough degrees of freedom, but so that the value of $\sigma_8$ stayed reasonable. To wit, our priors correspond to $\sigma_8 = 0.87 \pm 0.10$. For $P_1$ and $P_2$ we adopt a tight $10\%$ prior, since those bins are on scales larger than the survey window, and are in any case very weakly constrained due to the large cosmic variance. As is shown in Appendix~\ref{app:autocorr}, the $P_3$--$P_8$ priors remain reasonably broader than the posteriors. We have subsequently tested that using priors twice as wide only results in a 10\% widening of the final contours, and negligible shift on the posterior peaks.
For the shot-noise parameter, we adopt Gaussian priors which allows for a conservative 50\% effective variation of the galaxy number density $\Bar{n}$.

For the bias parameters, we employed Gaussian priors, using as reference the final NGC posterior results from~\cite{Ivanov:2019pdj} (henceforth ISZ20). We took their error bars, propagated the uncertainty in the power spectrum amplitude $A$ since they multiply all their bias parameters by $(A/A_{\rm Planck})^{1/2}$, symmetrized the results and multiplied them by a factor of four. We used the final results for the mean and standard deviation to construct our low-$z$ uncorrelated Gaussian priors. We also did a similar procedure for the counterterm parameters, but only multiplied the results by a factor of two. We thus allow a four times larger uncertainty of ISZ20 in each bias and twice as large for the counterterm parameters as our prior. This resulted in posteriors which were never strongly driven by the priors themselves. I.e., these priors are weakly informative.

For both $E$ and $L_A$ we take very large and uninformative uniform priors, centered around their respective $\Lambda$CDM value. We notice that current SN constraints on $L_A$ are already much tighter than this (for instance, taking the 16 SN in the redshift range $z\in(0.37,0.39)$ from the  Pantheon+~\cite{Brout:2022vxf} catalog produces an error around 3\%). However, their use would need the assumption of standard candles, while  in this paper we prefer to focus exclusively on clustering data.

We leave a more comprehensive study of the prior effects on the FreePower methodology for a future study.

\section{MCMC results}

We explore the 17-dimensional parameter space with the \texttt{Python} MCMC codes \texttt{emcee} \cite{Foreman-Mackey:2012any}\footnote{\url{https://emcee.readthedocs.io/en/stable} (last access: 22.03.2025)} and \texttt{ptemcee}~\cite{Vousden:2016eeu}.\footnote{\url{https://github.com/willvousden/ptemcee} (last access: 22.03.2025)}
We found that \texttt{emcee} exhibited very large autocorrelation times ($\sim 10000$) for most parameters, which led to a very slow convergence time even when using 256 walkers. We therefore resorted to the modified version \texttt{ptemcee}, which uses parallel tempering, for which the autocorrelation time was typically only $\sim 10$. All of our following results were obtained using \texttt{ptemcee}.

We ran \texttt{ptemcee} using 60 walkers and 6 different temperatures, initialized in a large volume in the allowed prior space. We ran, for each walker, 300k steps, for a total of 108 million chain points for each low-$z$ and high-$z$ samples. Due to the complexity of the 1-loop calculations and the amount of MCMC steps needed, the total computational time used for both redshift bin runs was large, around 50k CPU-hours. This heavy computation cost was the main reason for our choice of a small number of $k$-bins, limiting somewhat the dimensionality of our parameter space. We analyzed the results using both the \texttt{Python} package \texttt{getdist}~\cite{Lewis:2019xzd} and our own codes. We compared our results obtained with both \texttt{ptemcee} and \texttt{emcee} to confirm the results we found were robust. As we will discuss, we find a tension between our results and $\Lambda$CDM CMB results.

\begin{figure}
    \centering
    \includegraphics[width=.85\columnwidth]{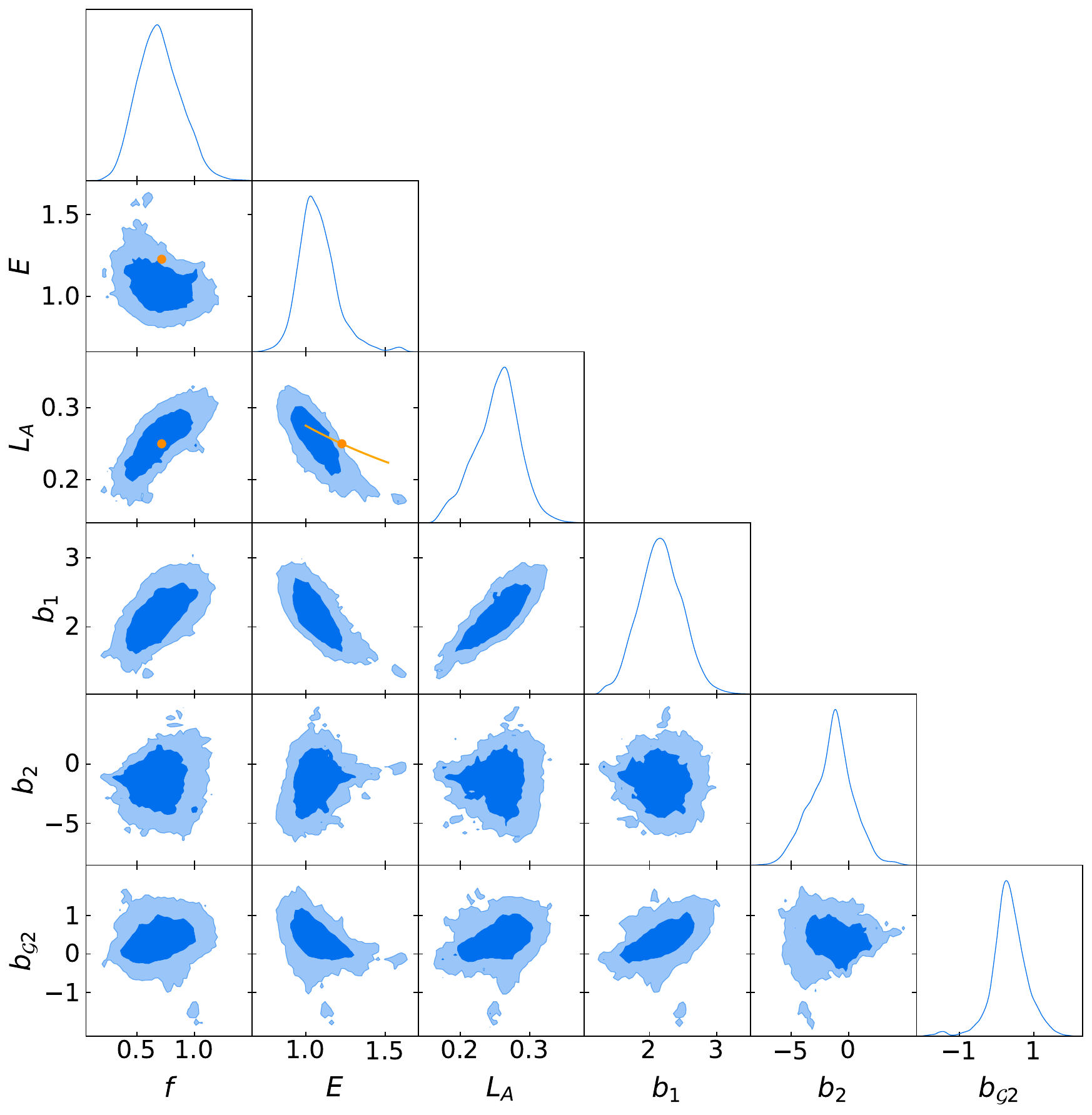}
    \caption{$68.3\%$ and $95.4\%$ highest-density interval corner plot for the cosmological and bias terms for the red low-$z$ sample ($z_{\rm eff}=0.38$). Orange dots mark the Planck 2018 $\Lambda$CDM values. The orange curve represents the $L_A$ vs.~$E$ relation for  $\Lambda$CDM for different values of $\Omega_{m0}$.
    }
    \label{fig:plot_lowz}
\end{figure}

\begin{figure}
    \centering
    \includegraphics[width=.85\columnwidth]{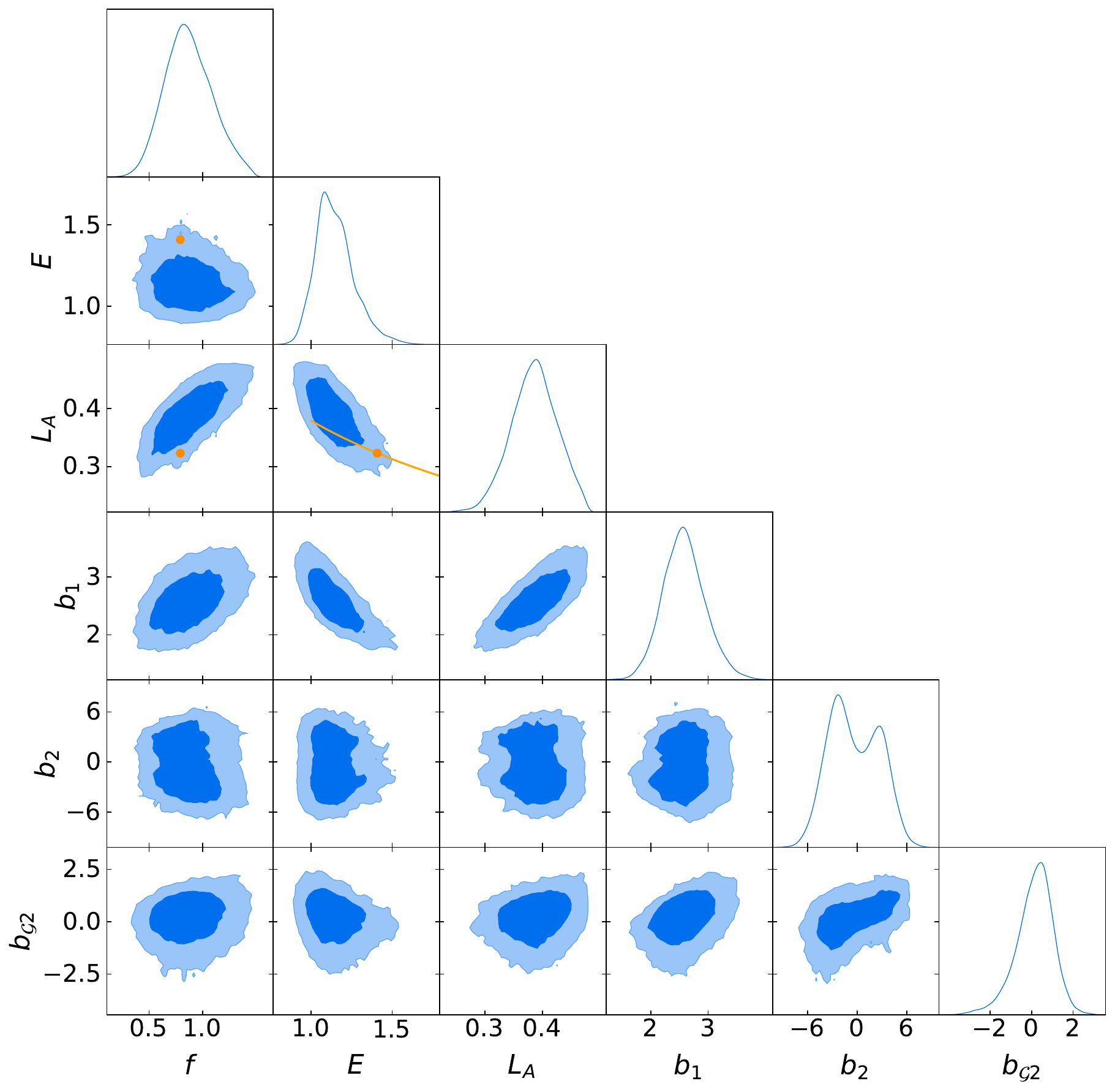}
    \caption{
    Same as Figure~\ref{fig:plot_lowz} for the high-$z$ sample ($z_{\rm eff}=0.61$).
    }
    \label{fig:plot_highz}
\end{figure}

Figures~\ref{fig:plot_lowz} and \ref{fig:plot_highz} show, for both redshift bins, the 68.3 and 95.4\% confidence intervals (CI) for $f$, $E$, $L_A$ and the bias parameters. These are our main results. The orange dots mark the Planck 2018 values assuming flat $\Lambda$CDM~\cite{Planck:2018vyg}, whereas the orange curves depict the $L_A(z)$ vs. $E(z)$ relation in flat $\Lambda$CDM when one varies the $\Omega_{m0}$ parameter.
Interestingly, as can be seen, both $z$-bins results for $E(z)$ are in small tension with flat Planck $\Lambda$CDM (and with any higher $\Omega_{m0}$). In particular, our results are at $1.7\sigma$ tension for $z=0.38$ and $2.1\sigma$ for $z=0.61$. For $z=0.61$ we also find $L_A$ to be larger than expected, away from the $\Lambda$CDM value at $1.8\sigma$.  On the other hand, all other parameters are mostly within 1$\sigma$ and always within  2$\sigma$  from  Planck 2018 $\Lambda$CDM  (except marginally for $P_7$ for high-$z$).

From the $P(k)$ linear bins we can also obtain the amplitude $\sigma_8$ by interpolating and integrating over a spherical window function. We find $\sigma_8$ to be $0.887 \pm 0.094$ ($0.799 \pm 0.083$) for the $z_{\rm eff} = 0.38$ ($0.61$) redshift bin. These results are, however, largely prior driven, since the priors on $P_i$ correspond to $\sigma_8 = 0.87 \pm 0.10$. We also remark that due to the small number of $k$-bins here considered, the value of $\sigma_8$ will inevitably depend on the chosen interpolation scheme.
We show the effect of the chosen interpolation scheme and number of $k$-bins on the obtained values for $\sigma_8$ in Appendix~\ref{app:interpolation}.

Table~\ref{tab:main-results} summarizes the marginalized constraints in each cosmological and bias parameter, whereas Table~\ref{tab:results-Pk} show the results for all 8 $k$-bins of $P(k)$. Both tables also show the best-fit Planck 2018 values assuming flat $\Lambda$CDM.

The correlations $\rho$ among the 17 parameters are in most cases small. The average off-diagonal absolute value of $\rho$ is around 0.17 for both redshift bins. We list here some particular values for $z_{\rm eff}=0.38$ (0.61): $\rho(f,E) = -0.32$ ($-0.12$), $\rho(E,L_A) = -0.80$ ($-0.69$), $\rho(E,b_1) = -0.75$ ($-0.76$), $\rho(L_A,b_1) = 0.83$ ($0.75$),  $\rho(c_0,c_2) = 0.65$ ($0.41$). There are some large correlations between individual $k$-bins of $P(k)$ and some bias parameters. But $f$, $E$ and $L_A$  show only small or moderate correlations with $P(k)$ and the bias parameters. The full correlation matrices for both bins are shown in Appendix~\ref{app:corrmat-convergence}.

\begin{table}
    \setlength\tabcolsep{4.2pt}
    \centering
    \begin{tabular}{lcccccc}
        \hline
        \hline
        \multirow{2}{*}{parameter} & $z_{\rm eff} = 0.38$ & $z_{\rm eff} = 0.38$ & $z_{\rm eff} = 0.61$ & $z_{\rm eff} = 0.61$ & $\Lambda$CDM & $\Lambda$CDM  \\
        & (HDI) & (ETI) & (HDI) & (ETI) & ($z=0.38$) & ($z=0.61$)\\
        \hline
        $f$                   &$0.67^{+0.20}_{-0.19}$    &[0.51, 0.90]  & $0.82^{+0.25}_{-0.20}$   &[0.65, 1.10]  & 0.716 & 0.792 \\
        $E$                   &$1.033^{+0.13}_{-0.081}$  &[0.97, 1.18]   & $1.085^{+0.16}_{-0.067}$ &[1.03, 1.26]   & 1.23  & 1.41  \\
        $L_A$                 &$0.264^{+0.026}_{-0.039}$ &[0.22, 0.28]  & $0.390^{+0.036}_{-0.046}$&[0.35, 0.43] & 0.250 & 0.323 \\
        $b_1$                 &$2.14^{+0.37}_{-0.29}$    &[1.8, 2.5]    & $2.54^{+0.37}_{-0.35}$   &[2.2, 2.9]   & --   & -- \\
        $b_2$                 &$-1.20^{+1.7}_{-1.9}$     &$[-3.4, 0.31]$  & $-2.37^{+6.1}_{-1.6}$    &$[-3.5, 3.1]$  & --   & -- \\
        $b_{\mathcal{G}_2}$   &$0.249^{+0.52}_{-0.32}$   &$[-0.045, 0.78]$& $0.42^{+0.69}_{-0.98}$   &$[-0.72, 1.00]$& --   & -- \\
        $c_0/[{\rm Mpc}/h]^2$ &$ 2^{+36}_{-45}$          &$[-61, 26]$     & $28^{+61}_{-56}$         &$[-28, 89]$    & --   & -- \\
        $c_2/[{\rm Mpc}/h]^2$ &$47^{+36}_{-46}$          &$[-2.0, 80]$    & $68^{+62}_{-60}$         &$[6.4, 130]$   & --   & -- \\
        $s$                   &$0.26^{+0.37}_{-0.56}$    &$[-0.31, 0.63]$ & $0.10^{+0.40}_{-0.45}$   &$[-0.33, 0.50]$& --   & -- \\
        \hline
        \hline
    \end{tabular}
    \caption{Main results for the two redshift bins. We show best-fits with 68.3\% highest-density interval (HDI), and also the 68.3\% equal-tailed intervals (ETI).  The last two columns show } the expected values for $\Lambda$CDM with Planck 2018 values~\cite{Planck:2018vyg}.
    \label{tab:main-results}
\end{table}

\begin{table}
    \setlength\tabcolsep{3.3pt}
    \centering
    \begin{tabular}{c|cccccccc}
    \hline
    \hline
        & $P_1$ & $P_2$ & $P_3$ & $P_4$ & $P_5$ & $P_6$ & $P_7$ & $P_8$ \\
        $k (h/$Mpc) & 0.0001 & 0.003 & 0.023 & 0.079 & 0.187 & 0.364 & 0.630 & 1.0
        \\ \hline
        $z_{\rm eff}=0.38$ & $437^{+34}_{-53}$ & $10500^{+850}_{-1100}$ & $21200^{+2800}_{-3400}$ & $7500^{+1000}_{-1200}$ & $2590^{+440}_{-470}$ & $1050^{+390}_{-360}$ & $270^{+250}_{-120}$ & $67^{+71}_{-56}$ \\
        $z_{\rm eff}=0.61$ & $427^{+45}_{-41}$ & $10300^{+940}_{-960}$ & $22100^{+2800}_{-3500}$ & $6800^{+1500}_{-940}$ & $1830^{+410}_{-480}$ & $650^{+360}_{-180}$ & $610^{+260}_{-270}$ & $49^{+150}_{-34}$ \\
        $\Lambda$CDM ref. & 427 & 9980 & 23300 & 8560 & 2220 & 618 & 196 & 69.5 \\
    \hline
    \hline
    \end{tabular}
    \caption{Best-fit and 68.3\% HDI for each $k$-bin of the linear power spectrum (at $z=0$), in units of $({\rm Mpc}/h)^3$. The $k$ values represent the center of each $k$-bin. The model-independent constraints on $P_i(z)$ were converted to $z=0$ values for comparison using the $\Lambda$CDM growth factor $G(z)=0.817 \hspace{2mm}(0.728)$ for $z=0.38 \hspace{2mm} (0.61)$. The last row reports the corresponding Planck 2018 $\Lambda$CDM values as reference values.
    \label{tab:results-Pk}}
\end{table}

We tested the convergence of our MCMC using primarily the auto-correlation time ($\tau$) estimates~\cite{Foreman-Mackey:2012any}. A large ratio between the number of steps and $\tau$ indicate that the chain ran for long enough to have a large effective number of independent steps. $\tau$ should converge to an asymptotic value for high enough number of steps. We found that using \texttt{ptemcee} $\sim 1000$ steps were necessary to reach convergence, while using \texttt{emcee} one needed almost a million steps. We also performed the simple Gelman-Rubin test~\cite{Gelman:1992zz}. Both results indicate complete convergence of the chains.  More details are provided in Appendix~\ref{app:corrmat-convergence}.

\section{Comparison with previous analyses of BOSS data}

In the following, the results obtained in this study are compared with the constraints previously obtained in \emph{model-dependent} analyses. It is important to note that the approaches applied in these analyses and the model-independent approach used here are quite different. An analysis of the nonlinear galaxy power spectrum obtained from BOSS under the model assumption of $\Lambda$CDM with varied neutrino masses was carried out in ISZ20. Another analysis of the BOSS data including information from power spectrum multipoles, the bispectrum monopole, the real-space power spectrum and the reconstructed power spectrum was carried out in~\cite{Philcox:2021kcw} (henceforth, PI22). For each bias parameter, the authors report four results. Here we focus on the comparisons without the bispectrum and with a free spectral slope $n_s$.

\begin{table}
    \setlength\tabcolsep{3.8pt}
    \centering
    \begin{tabular}{lccc|ccc}
    \hline\hline
    par.               & this work              & ISZ20                 & PI22                   & this work               & ISZ20                  & PI22                 \\
    \hline
    & \multicolumn{3}{c|}{$z_{\rm eff} = 0.38$} & \multicolumn{3}{c}{$z_{\rm eff} = 0.61$} \\
    \hline
    $f$                 & $0.70^{+0.20}_{-0.19}$     & $0.735\pm 0.084$          & ---                    & $0.87^{+0.23}_{-0.22}$         & $0.750\pm 0.085$ & ---   \\
    $E$                 & $1.08 ^{+0.10}_{-0.11}$    & $1.138 \pm 0.049$         & ---                    & $1.15 \pm 0.11$        & $1.386\pm 0.076$ & ---   \\
    $L_A$                 & $0.25^{+0.03}_{-0.04}$    & $0.242 \pm 0.006$         & ---                    & $0.38 \pm 0.04$        & $0.309 \pm 0.012$ & ---   \\
    $b_1$               & $2.16\pm 0.33$     & $1.67^{+0.16}_{-0.20}$    & $2.21\pm0.14$          & $2.57\pm 0.36$           & $2.28^{+0.22}_{-0.30}$ & $2.33\pm0.15$          \\
    $b_2$               & $-1.48^{+1.70}_{-1.86}$  & $-2.3^{+0.5}_{-0.9}$  & $-0.51^{+0.79}_{-0.97}$ & $-0.34^{+3.45}_{-3.14}$            & $-2.7^{+0.5}_{-2.1}$   & $-1.09^{+0.84}_{-1.00}$ \\
    $b_{\mathcal{G}_2}$ & $0.34^{+0.44}_{-0.39}$ & $0.11^{+0.28}_{-0.40}$ & $-0.38\pm0.37$         & $0.12\pm 0.85$ & $0.59^{+0.26}_{-0.08}$ & $-0.19\pm0.43$         \\
    \hline\hline
    \end{tabular}
    \caption{Comparison with the literature. Here we quote mean values and 68.3\% equal-tailed intervals, instead of highest-density ones, since those are the ones reported in~\cite{Philcox:2021kcw,Ivanov:2019pdj} (we show the ISZ20 case with an $\omega_b$ prior).
    }
    \label{tab:comp-all}
\end{table}

A comparison of the inferred cosmological parameters with these works is given in Table~\ref{tab:comp-all}. In this table we quote mean values and 68\% percentiles using equal-tailed statistics, instead of highest-density values, since those are the ones reported in ISZ20 and PI22. For the low-$z$ survey we find $f=0.7\pm 0.2$ and $E=1.08\pm 0.10$, in reasonable agreement with $f=0.74\pm 0.08$, $E=1.14\pm 0.05$ from ISZ20. FreePower measures directly $f(z)$, while conventional analysis measure instead $f(z)\sigma_8(z)$. We therefore convert the latter using the values of $\sigma_8$ and $\Omega_{m0}$ obtained in ISZ20 and  propagate the uncertainty  in the integral for the growth factor.

\begin{figure}
\centering
\includegraphics[width=.65\columnwidth]{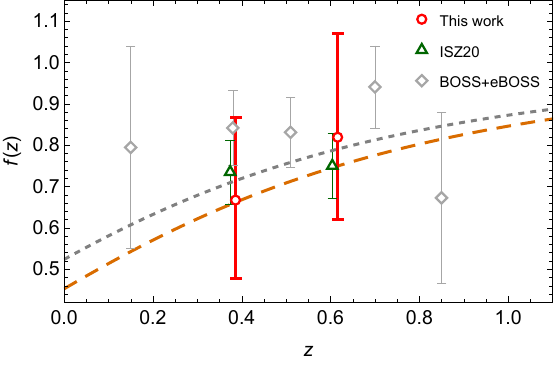}
    \caption{Comparison of our model-independent $f(z)$ results with those obtained assuming $\Lambda$CDM in ISZ20 and with BOSS+eBOSS galaxies~\cite{eBOSS:2020yzd}.  FreePower measures directly $f(z)$ but conventional analysis measure instead $f(z)\sigma_8(z)$, so we convert the latter using the values of $\sigma_8$ and $\Omega_{m0}$ obtained in ISZ20. The best fit Planck 2018 curves are depicted in dashed gray ($\Lambda$CDM) or long-dashed orange ($\gamma\Lambda$CDM~\cite{Nguyen:2023fip} -- see text).
    }
\label{fig:plot_f}
\end{figure}

In Figure~\ref{fig:plot_f} we compare our results for $f(z)$ with ISZ20 and another previous analysis of BOSS and eBOSS galaxies~\cite{eBOSS:2020yzd} which assume $\Lambda$CDM. As can be seen, our model-independent results favor smaller values for $f$, which is in qualitative agreement with a recent analysis of the  Planck 2018 data in a flat $\gamma\Lambda$CDM model, in which the growth-rate index $\gamma$, defined as $f\simeq \Omega_m(z)^\gamma$, was left free to vary~\cite{Nguyen:2023fip}.

We note that also the bias $b_1,b_2,b_{\cal G}$ agree to within 1$\sigma$ with ISZ20 and PI22. For the high-$z$ survey, we find $f=0.87\pm 0.22, E=1.15\pm 0.11$, again to within 1-1.5$\sigma$ from ISZ20 ($f=0.75\pm 0.08, E=1.38\pm0.07$). The bias parameters are in 1$\sigma$ agreement as well.

As expected, our errors are always significantly larger than in model-dependent analyses. For instance, the errors on $f$, $E$ and $L_A$ are roughly twice as large.

We also compare our results to the BOSS results \cite{2017MNRAS.470.2617A}. They find for $f\sigma_8$ at $z_{\rm eff} =0.38$ $0.502\pm0.065$ and $0.497\pm0.063$ for full-shape and joint (BAO + FS) measurements, which agrees very well with our result of $0.49\pm0.16$. At $z_{\rm eff}=0.61$, they find $0.419\pm 0.045$ and $0.436\pm 0.043$, which is also within $1\sigma$ of our result $0.48^{+0.15}_{-0.12}.$

\section{Comparison of distance measurements}

\begin{figure}
\centering
\includegraphics[width=.65\columnwidth]{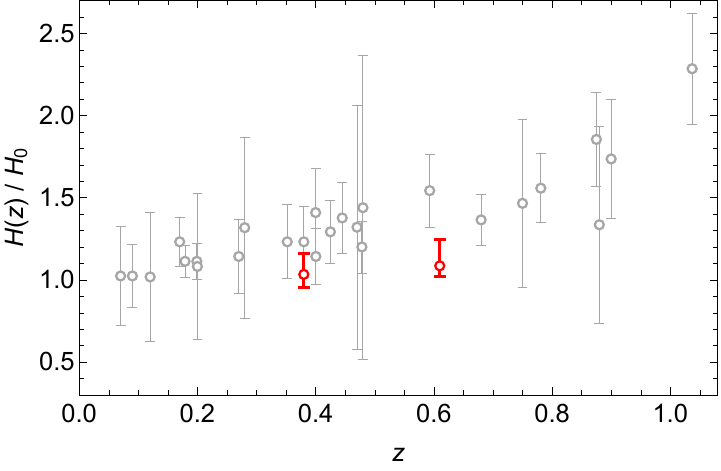}
    \caption{Comparison of $E(z)$ results using cosmic chronometers (CC, in gray) and our FreePower result (in red).  For the CC data we use the compilation in~\cite{Moresco:2022phi}, limited to $z<1.2$, and use the $H_0$ value obtained from CC data itself with a Gaussian Process extrapolation~\cite{Gomez-Valent:2018hwc}.
    }
\label{fig:plot_CC}
\end{figure}

Few methods are capable of direct measurements of the expansion rate (or of the corresponding Hubble distance) without an underlying assumption of the cosmological model. One such method relies on the use of the so-called cosmic chronometers (CC), which are based on modelling passively evolving galaxies. In Figure~\ref{fig:plot_CC} we compare our results for $E(z)$ with a recently compiled catalog of CC~\cite{Moresco:2022phi}. For clarity, we only show the CC for $z<1.2$, but this CC dataset extends until $z\simeq2$. Since CC measure $H(z)$ instead of $E(z)$, we convert their measurements  using the value of  $H_0 = 67.42 \pm 4.75$, obtained from CC data itself with a Gaussian Process extrapolation~\cite{Gomez-Valent:2018hwc}. As can be seen, our results are very competitive, and compensate the smaller number of data points with smaller error bars.

Recently, the first release of DESI data \cite{DESI:2024mwx} was analyzed and constraints have been put forward for $E$ and $L_A$ at seven redshift bins from 0.3 to 2.33 (effective values). To compare directly to DESI tables and figures, we first convert our $E, \,L_A$ constraints into the distance parameters $D_H, \, D_M$ as
\begin{align}
    \frac{D_{M}}{r_{d}}&=\frac{L_{A}(1+z)}{H_{0}r_{d}},\quad\frac{D_{H}}{r_{d}}=\frac{1}{H_{0}r_{d} E} \,,
\end{align}
and then into the basis employed by the DESI collaboration for some of their figures:
\begin{align}
    \frac{D_{V}}{r_{d}}&=\frac{1}{r_d}\big(zD_{M}^{2}D_{H}\big)^{1/3},\quad F_{AP}=\frac{D_{M}}{D_{H}} \,,
\end{align}
where $r_d$ is the sound horizon measured on the CMB.  When we tabulate our results for $D_M/r_d,D_H/r_d,D_V/r_d$  we  adopt for $H_0 r_d$ the best fit  $\Lambda$CDM value of the DESI paper, namely
\begin{equation}
    r_d h=\frac{r_dH_0}{100\,({\rm km/sec/Mpc})}=101.8\pm 1.3  \,{\rm Mpc}\,.
\end{equation}
This is however quite different from  Planck result
\begin{equation}
    r_d h=\frac{r_dH_0}{100\,({\rm km/sec/Mpc})}=98.82\pm 0.82  \,{\rm Mpc} \,.
\end{equation}
so the comparison of the $H_0 r_d$-dependent quantities might be misleading.  $F_{AP}$, on the other hand, is independent of $H_0 r_d$. We also include  the measurements  obtained in ISZ20 with the $\alpha$-analysis, which is a more  model-independent analysis of the BOSS data. In this section, we always use the mean values and symmetric 68\% percentiles instead of highest-density values (for $L_A$, both coincide).

\begin{figure}
\centering
\includegraphics[width=.75\columnwidth]{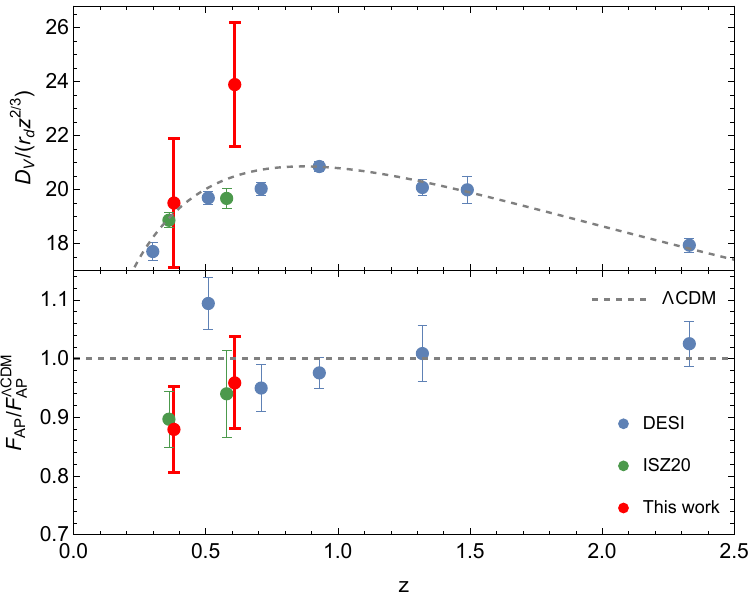}
    \caption{
    \emph{Top:} Comparison of $D_V/r_d$ (rescaled by $z^{2/3}$ to match the DESI prescription)  of DESI BAO (blue dots) and ISZ20 (green dots, slightly displaced for clarity) with our results (red dots).  The dashed line is the best fit Planck 2018  $\Lambda$CDM curve. \emph{Bottom:} same for $F_{AP}$ divided by the Planck 2018 $\Lambda$CDM $F_{AP}$ (note that in Ref.~\cite{DESI:2024mwx} $F_{AP}$ was normalized to the similar, but not identical, DESI $\Lambda$CDM best fit).
    }
\label{fig:plot_desidv}
\end{figure}

The results are illustrated  in Figure~\ref{fig:plot_desidv}, which is to be compared with Figure~1 of~\cite{DESI:2024mwx}. Since for both $z$-bins we find  $E$  below,  and $L_A$ above, Planck $\Lambda$CDM value, we obtain  larger $D_H,D_M$ and correspondingly  a substantially larger $D_V/r_d$ than $\Lambda$CDM, and also large associated errors. However, we do not attach particular importance to this quantity, since, as already mentioned, it depends on the sound horizon $r_d$ that we cannot measure in our approach.
For $F_{AP}$ the tension with the fiducial $\Lambda$CDM result at low-$z$ is just below 2$\sigma$. Our $F_{AP}$ results agree quite well with ISZ20. Interestingly, the deviation of $F_{AP}$ from $\Lambda$CDM we find at both redshift bins goes in the opposite direction with respect to the one detected by DESI in the intermediate sample at $z=0.51$. Table~\ref{tab:distances} summarizes all the distances.

\begin{table}
\centering
    \begin{tabular}{rlcccccc}
    \hline\hline
    \multicolumn{2}{c}{data} &  $z$              & $D_H/r_d$             & $D_M/r_d$               & $\rho$  &          $D_V/(r_d z^{2/3})$ & \raisebox{0pt}[12pt][5pt]{}$F_{AP}/F^{\Lambda CDM}_{AP}$   \\
    \hline
     && 0.30 & -- & -- & -- & 17.70$\pm$0.33 & -- \\
    && 0.51 & 21.0$\pm$0.6 & 13.62$\pm$0.25 & -0.445 & 19.69$\pm$0.23 & 1.10$\pm$0.04 \\
    && 0.71 & 20.1$\pm$0.6 & 16.85$\pm$0.32 & -0.42 & 20.02$\pm$0.25 & 0.95$\pm$0.04 \\
   \multicolumn{2}{c}{\raisebox{0pt}[0pt][0pt]{\rotatebox[origin=c]{90}{DESI}}} & 0.93 & 17.88$\pm$0.35 & 21.71$\pm$0.28 & -0.389 & 20.85$\pm$0.18 & 0.98$\pm$0.03 \\
    && 1.32 & 13.8$\pm$0.4 & 27.8$\pm$0.7 & -0.444 & 20.07$\pm$0.30 & 1.01$\pm$0.05 \\
    && 1.49 & -- & -- & -- & 20.00$\pm$0.50 & -- \\
    && 2.33 & 8.52$\pm$0.17 & 39.7$\pm$0.9 & -0.477 & 17.93$\pm$0.25 & 1.03$\pm$0.04 \\
    \hline
    \raisebox{0pt}[5pt][5pt]&
& 0.38 & 26.2$\pm$1.3 & 9.9$\pm$1.3 & -- & 18.86$\pm$0.28 & 0.90$\pm$0.05\\
    \multicolumn{2}{c}{\raisebox{6pt}[5pt][5pt]{\rotatebox[origin=c]{90}{ISZ20}}} & 0.61 & 21.5$\pm$1.3 & 14.1$\pm$1.3 & -- & 19.70$\pm$0.4 & 0.94$\pm$0.07
   \\
    \hline
    & & 0.38 & 27.3$\pm$2.8 & 10.2$\pm$1.4 & 0.808 & 19.5$\pm$2.4 & 0.88$\pm$0.07\\
    \raisebox{0pt}[0pt][0pt]{\rotatebox[origin=c]{90}{\;\;\;this}} &\raisebox{0pt}[0pt][0pt]{\!\!\!\!\!\rotatebox[origin=c]{90}{\;\;\;work}} & 0.61 & 25.6$\pm$2.6 & 18.0$\pm$1.9 & 0.688 & 23.9$\pm$2.4 & 0.96$\pm$0.08\\
    \hline\hline
\end{tabular}
\caption{Distance measurements in DESI, ISZ20, and this work.
}\label{tab:distances}
\end{table}

\section{Conclusions and outlook}

The BOSS DR12 datasets at effective redshifts $0.38$ and $0.61$ for the NGC survey have been analyzed following, for the first time, the model-independent FreePower approach. The linear matter power spectrum has been parametrized in several $k$-bins and varied in the MCMC, along with the dimensionless expansion rate, the dimensionless angular diameter distance, the growth rate, and many other parameters that model the non-linear correction. It was shown that consistent results for various cosmological and non-cosmological parameters can be obtained from the observational data with only weak assumptions about the underlying cosmological model in terms of background expansion and power spectrum shape.  The results obtained showcase the potential of model-independent analyses of galaxy surveys.

The main findings of this study are summarized below:

\begin{itemize}
    \item We were able to constrain both non-cosmological parameters (bias, shot-noise and counterterm parameters), and cosmological parameters in a  model-independent way using LSS surveys. In particular, we measured $f(z),E(z)$ and $L_A(z)$ at both BOSS redshift bins (Table~\ref{tab:main-results}).

    \item Our constraints are consistent with the ones from previous model-dependent analyses \cite{Ivanov:2019pdj,Philcox:2021kcw}.
    The uncertainties increase in the model-independent analysis, as expected, due to it having fewer assumptions.

    \item
    We find a slight tension at around $2\sigma$ with $\Lambda$CDM results for $E$ in both redshift bins. For $L_A$, it is in good agreement for $z_{\rm eff} = 0.38$ but 1.7$\sigma$ larger than expected from $\Lambda$CDM for $z_{\rm eff} = 0.61$. In both bins the growth rate $f$ agrees with $\Lambda$CDM. For the bias parameters we find good agreement with a previous 1-loop analysis in both bins.

    \item We find a value of $F_{AP}$ in agreement with previous BOSS analyses. Our $F_{AP}$ values are both below Planck $\Lambda$CDM and around 2.5$\sigma$ away  from  the recent DESI estimate for the adjacent $z=0.51$ redshift bin.

\end{itemize}

This work demonstrates the possibility of implementing  a theoretical model for non-linear structure formation together with the model-independent galaxy survey analysis approach. It paves the way for model-independent analyses of future observational data of the large-scale structure of the Universe.
The estimates of $f, E, L_A$ can be employed to answer fundamental questions like the spatial curvature~\cite{Amendola:2024gkz} or the cosmological Poisson equation~\cite{Zheng:2023yco} independently of the initial conditions (the power spectrum shape) and of the late-time evolution (expansion and growth rates).

There are of course many ways to extend our results in the future. In terms of data, the hexadecapole can be included in the analysis next to monopole and quadrupole. One can also include the bispectrum~\cite{Philcox:2022frc}, which is of the same order as the one-loop power spectrum in the non-linear perturbation theory~\cite{Philcox:2021kcw}.  The BOSS SGC data can also be included, which would increase the effective volume of the sample by around 37\%~\cite{Ivanov:2019pdj}. In terms of methodology, the growth rate $f(z)$ could be generalized to be scale-dependent, and/or one could try using a larger number of $k$-bins. One could also use more informative priors for the dimensionless distance $L_A$ taken from external datasets like SN Ia.
Another possibility is to perform a combined analysis of both redshift bins, since in FreePower they share the same linear $P(k)$ parameters. In terms of theory, our analysis can be expanded by including deviations from the EdS kernels \cite{2008JCAP...10..036P,Bose:2018orj,Piga:2022mge}. It can also be extended to include the peculiar-velocity power spectrum~\cite{Amendola:2019lvy,Howlett:2019bky,Quartin:2021dmr}. A more thorough investigation of the effects on accuracy of different choices of $k_{\rm max}$ in each redshift bin is also desirable.
We note that a similar model-independent analysis can be carried out to investigate possible primordial non-Gaussianity (PNG)~\cite{Cabass:2022wjy}. We plan to pursue these lines of investigation in future work.

\section*{Acknowledgements}

We thank Sandro Vitenti for help with the MCMC convergence analysis, Louis Legrand and Vivian Miranda for tips regarding MCMC samplers, Massimo Pietroni and Marco Marinucci for several discussions and Eoin Colgáin for comments on our manuscript. MQ is supported by the Brazilian research agencies Fundação Carlos Chagas Filho de Amparo à Pesquisa do Estado do Rio de Janeiro (FAPERJ) project E-26/201.237/2022, CNPq (Conselho Nacional de Desenvolvimento Científico e Tecnológico) and CAPES. LA acknowledges support by the Deutsche Forschungsgemeinschaft (DFG, German Research Foundation) under Germany's Excellence Strategy EXC 2181/1 - 390900948 (the Heidelberg STRUCTURES Excellence Cluster) and under  project  456622116. We acknowledge the use of the computational resources of the joint CHE / Milliways cluster, supported by a FAPERJ grant E-26/210.130/2023. This study was financed in part by the Coordenação de Aperfeiçoamento de Pessoal de Nível Superior - Brasil (CAPES) - Finance Code 001. We acknowledge support from the CAPES-DAAD bilateral project  ``Data Analysis and Model Testing in the Era of Precision Cosmology''. %

\section*{Appendix}

\appendix

\section{Correlations and convergence}\label{app:corrmat-convergence}

\begin{figure}
\centering
\includegraphics[height=.43\columnwidth]{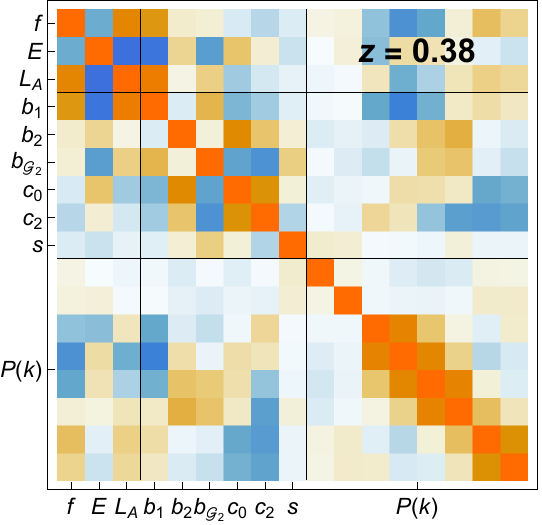}
\includegraphics[height=.43\columnwidth]{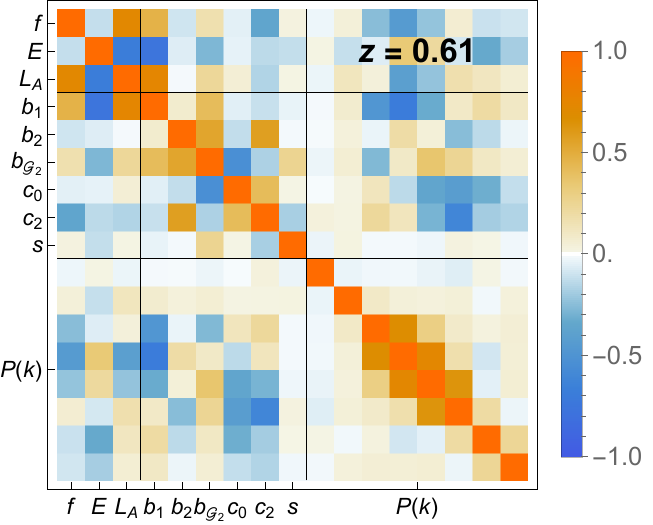}\\
\vspace{-0.2cm}
\caption{Correlation matrix for all 17 parameters in both redshift bins. The structure is similar in both bins; at low-$z$ the correlations among the bias parameters is slightly larger.
}
\label{fig:corrmat}
\end{figure}

The correlation matrices among all 17 parameters are depicted in Figure~\ref{fig:corrmat}, for both redshift bins. As discussed in the text, most terms are small. The bias parameters have stronger correlations among themselves for the lower redshift bin. The different $P(k)$ bins have strong positive correlations with their immediate neighbors, except for  $P_1$ and $P_2$, for which tight priors were used. The data-driven cosmological parameters $f$, $E$ and $L_A$ are correlated among themselves and $b_1$. $E$ is also correlated with $b_{\mathcal{G}_2}$.

\begin{figure}
\centering
\includegraphics[width=.4\columnwidth]{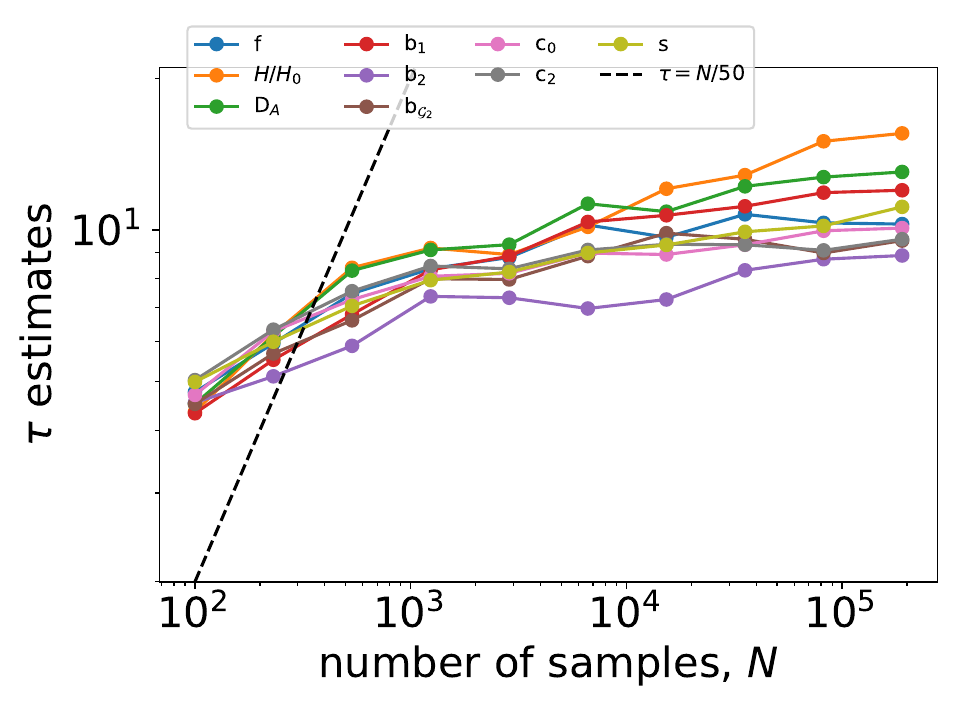}
\includegraphics[width=.4\columnwidth]{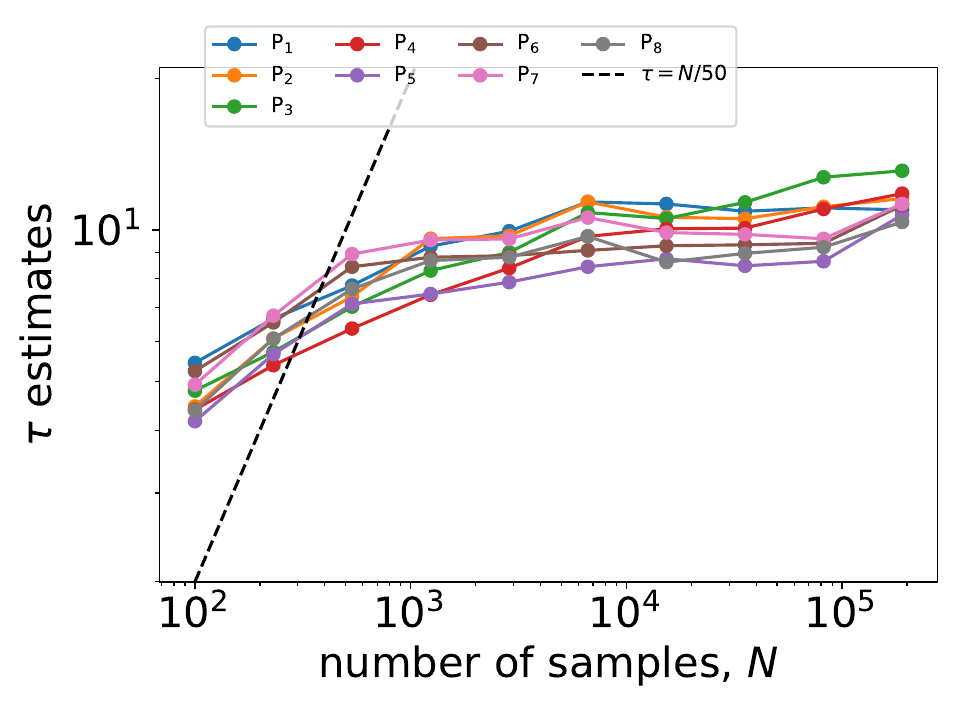}\\
\vspace{-0.2cm}
\includegraphics[width=.4\columnwidth]{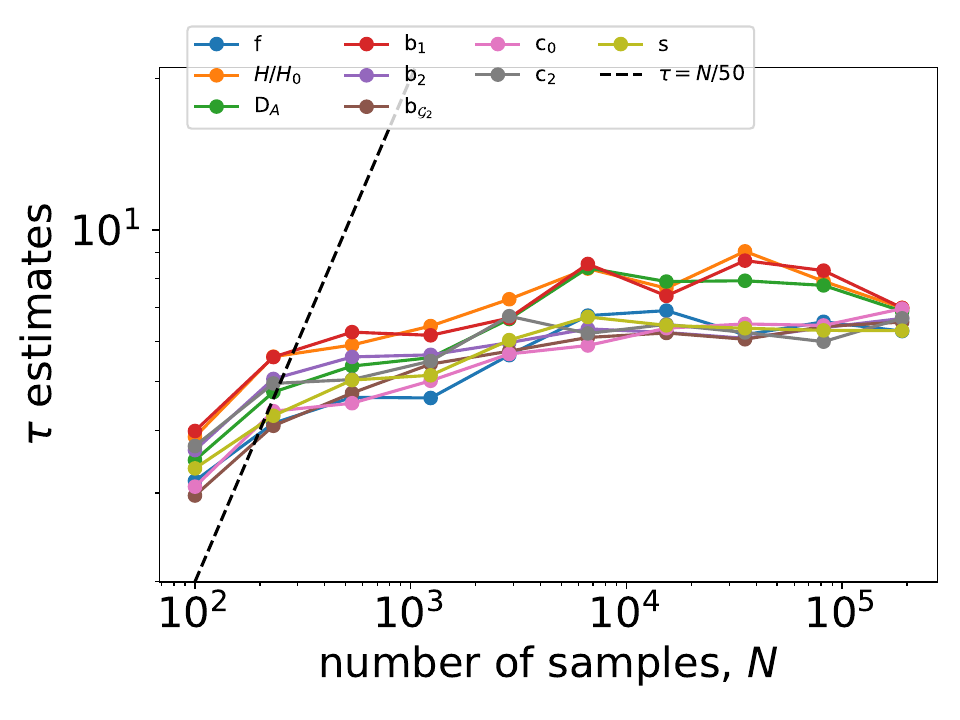}
\includegraphics[width=.4\columnwidth]{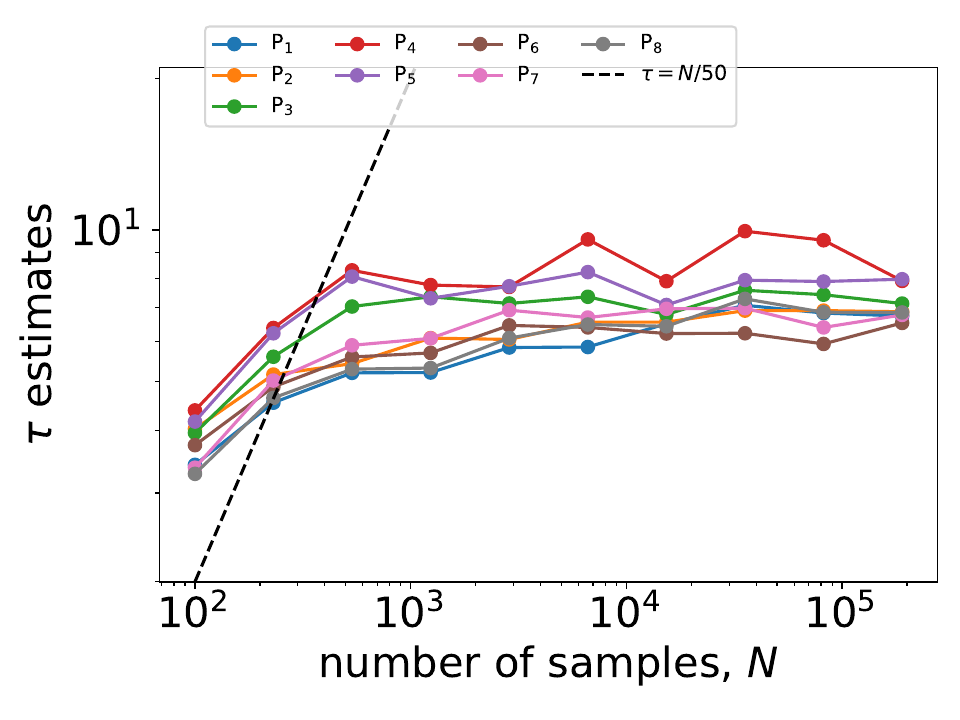}
\caption{Integrated autocorrelation times as a function of the number of steps in each of the 60 walkers. \emph{Top:} $z_{\rm eff}=0.38$; \emph{bottom:} $z_{\rm eff}=0.61$.
}
\label{fig:autocorr}
\end{figure}

We also display the integrated autocorrelation times ($\tau$) in Figure~\ref{fig:autocorr}. We compute $\tau$ for each walker individually and these estimates are averaged afterwards~\cite{Foreman-Mackey:2012any}. The MCMC shows clear signs of convergence as the estimates for the $\tau$ crosses the $N/50$ line, where $N$ is the number of MCMC steps. Note that $\tau\sim10$ for all parameters,
but we ran many more than 500 steps to get smoother 2D contours. We do not perform any thinning of the chains.

Regarding the Gelman-Rubin test, for all parameters  we get $(\hat{R})^{(1/2)} - 1 < 0.0001$ for both low and high-$z$ runs. We note that $(\hat{R})^{(1/2)}$ is not the most reliable test, specially using ensemble samplers like \texttt{ptemcee}, since in this approach the walkers are not independent. Nevertheless, since the results are really very close to unity, they also point to a very good convergence of all chains.

\section{Full corner plots}\label{app:autocorr}

\begin{figure}
    \centering
    \includegraphics[width=0.65\columnwidth]{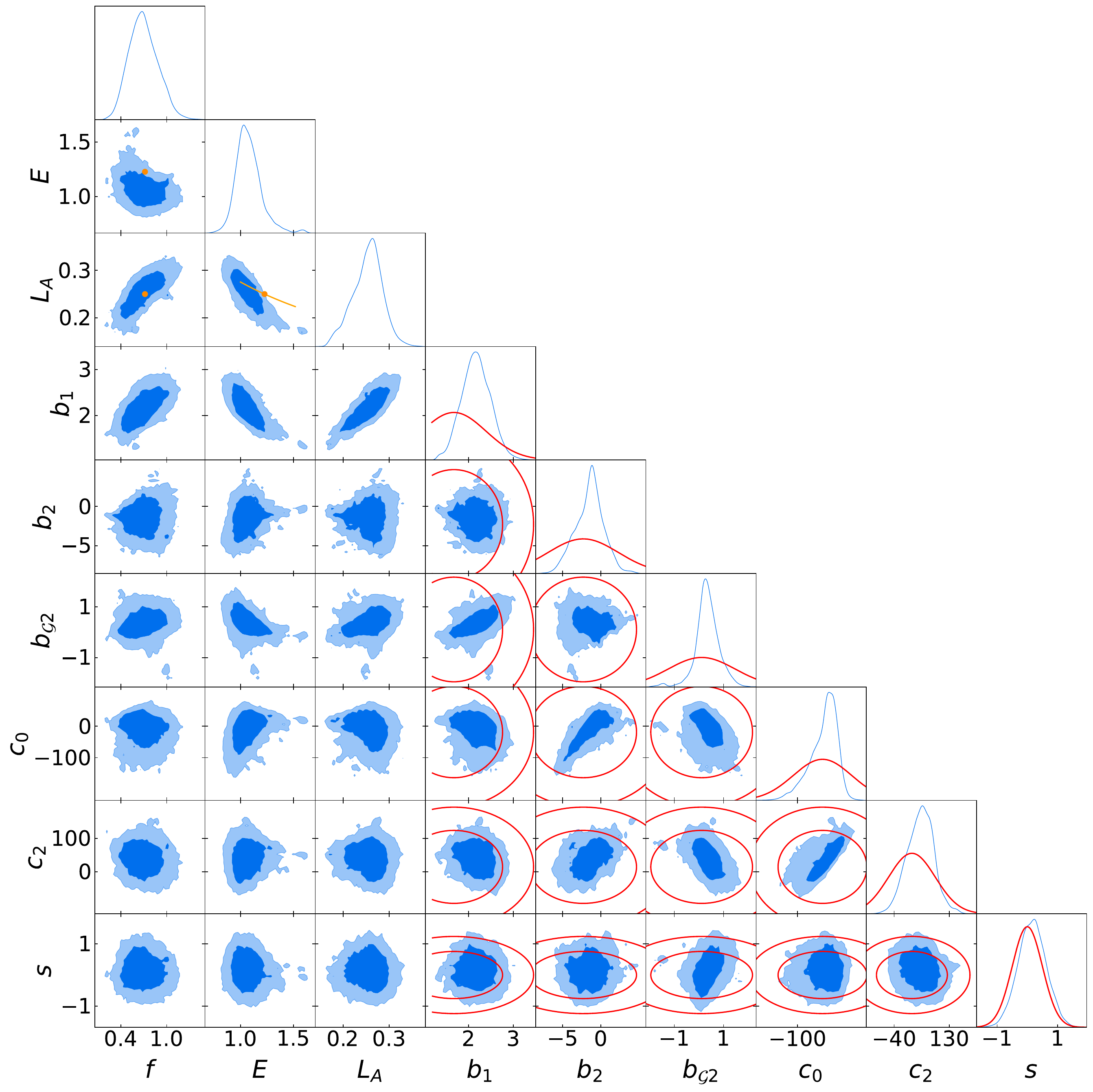}
    \caption{Similar to Figure~\ref{fig:plot_lowz} for $z_{\rm eff}=0.38$, but including counterterms and shot-noise parameters. The priors are depicted in red; the green points mark the Planck  $\Lambda$CDM values.
    }
    \label{fig:plot_lowz_full_1}
\end{figure}

\begin{figure}
    \centering
    \includegraphics[width=0.65\columnwidth]{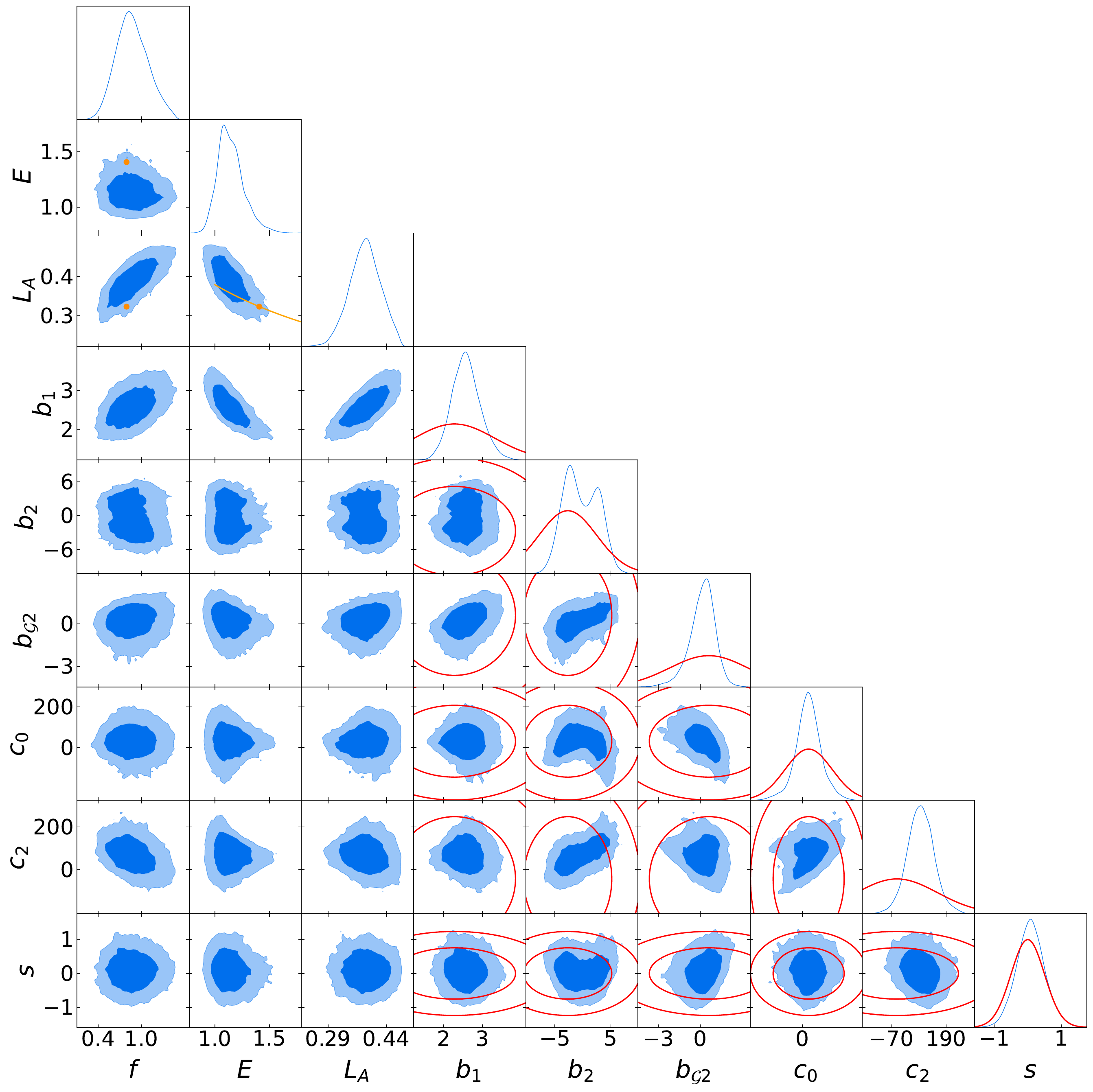}
    \caption{Same as Figure~\ref{fig:plot_lowz_full_1} for $z_{\rm eff}=0.61$.}
    \label{fig:plot_highz_full_1}
\end{figure}

\begin{figure}
    \centering
    \includegraphics[width=0.65\columnwidth]{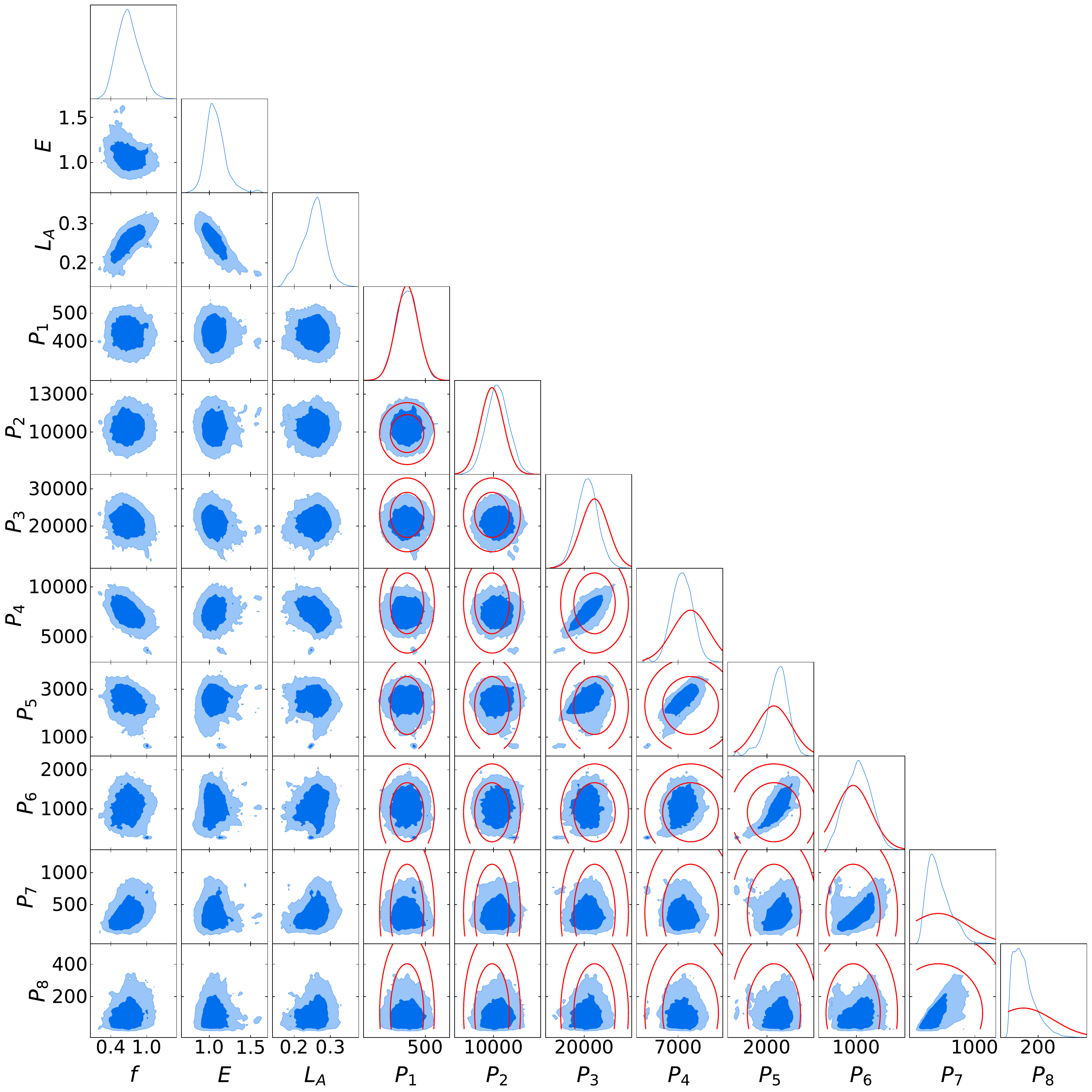}
    \caption{
    Same as Figure~\ref{fig:plot_lowz_full_1} for $z_{\rm eff}=0.38$ but for the parameters describing the cosmology and the linear matter power spectrum.}
    \label{fig:plot_lowz_full_2}
\end{figure}

\begin{figure}
    \centering
    \includegraphics[width=0.65\columnwidth]{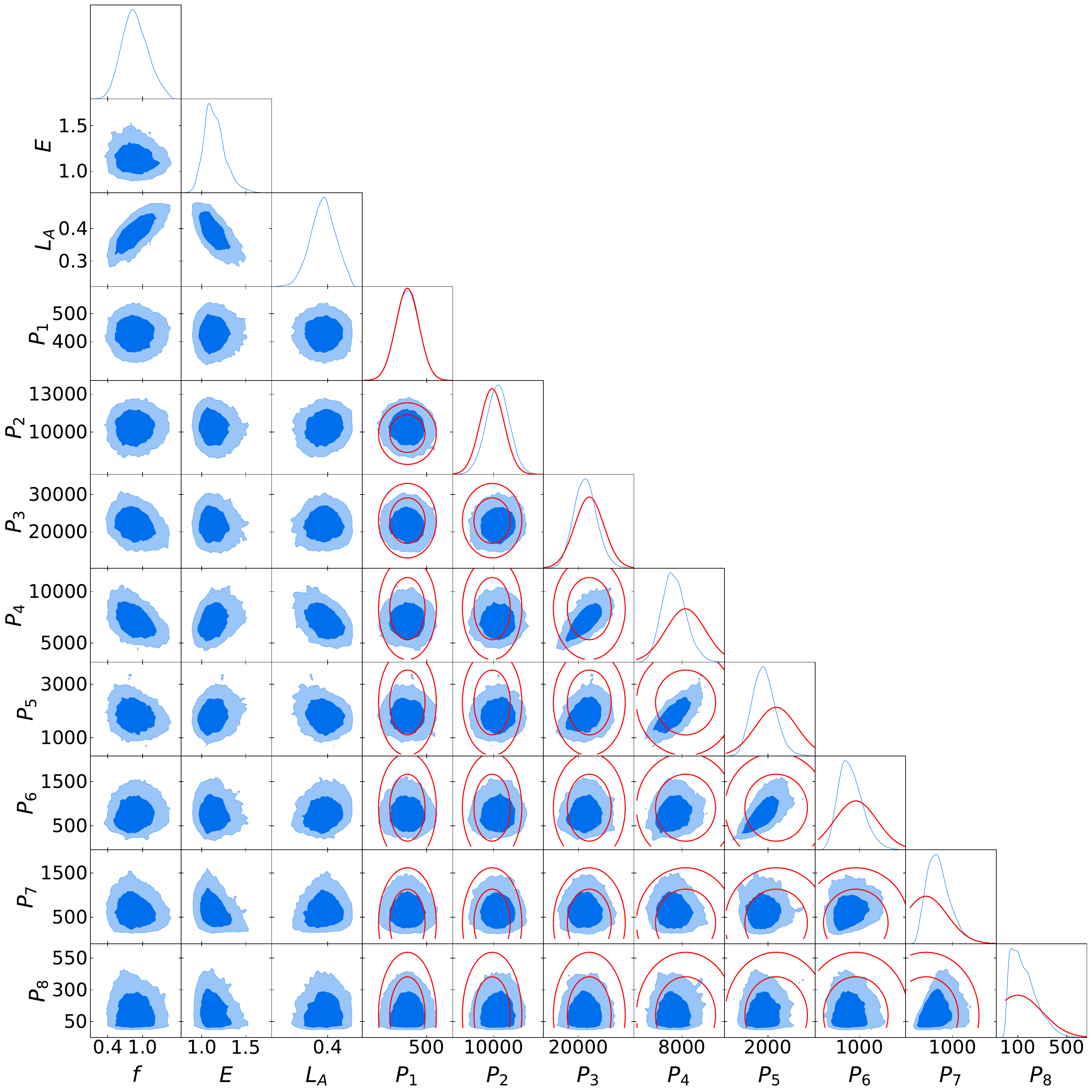}
    \caption{
    Same as Figure~\ref{fig:plot_lowz_full_2} for $z_{\rm eff}=0.61$.}
    \label{fig:plot_highz_full_2}
\end{figure}

Here we present larger triangle plots, which include more parameters and also the prior ranges. Figures~\ref{fig:plot_lowz_full_1} and~\ref{fig:plot_highz_full_1} are extensions of Figure~\ref{fig:plot_lowz} and~\ref{fig:plot_highz}, including also the counterterm and shot-noise parameters. Figures~\ref{fig:plot_lowz_full_2} and \ref{fig:plot_highz_full_2} instead show the parameters depicting the power spectrum bins, together with the growth function and expansion rate, which are the two data-driven cosmological parameters.

\section{One-loop galaxy power spectrum}\label{app:one-loop}

The galaxy power spectrum includes model parameters describing the galaxy bias \cite{Simon:2022lde}, counterterms, and the shot noise through the dimensionless shot noise parameter
\begin{equation}\label{eq:shot}
    s=P_{\rm shot}\Bar{n}-1\,,
\end{equation}
where $\Bar{n}\approx 3\times 10^{-4} (h/{\rm Mpc})^3$~\cite{Ivanov:2019pdj} is the number density of the galaxies in BOSS. The parameters are in the so-called EC basis~\cite{Simon:2022lde}.

The one-loop corrections for the galaxy density field power spectrum have studied in many works (see e.g.~the  review \cite{Bernardeau:2001qr}) and, when applied to tracers, take the form \cite{Ivanov:2019pdj}
\begin{equation}
    P_{11}(k,\mu,z)=\big[Z_1^{\rm sym}(\mathbf{k})\big]^2P_{\rm L}(k,z),
\end{equation}
\begin{equation}
    P_{22}(k,\mu,z)=2\int {\rm d}\mathbf{q}
    \big[Z_2^{\rm sym}(\mathbf{q},\mathbf{k}-\mathbf{q})\big]^2P_{\rm L}(q,z)P_{\rm L}(\abs{\mathbf{k}-\mathbf{q}},z)\
\end{equation}
and
\begin{equation}
    P_{13}(k,\mu,z) = 6 Z_1^{\rm sym}(\mathbf{k})\int {\rm d}\mathbf{q} \, Z_3^{\rm sym}(\mathbf{k},\mathbf{q},-\mathbf{q})P_{\rm L}(k,z)P_{\rm L}(q,z),
\end{equation}
where $Z_1^{\rm sym}$, $Z_2^{\rm sym}$, and $Z_3^{\rm sym}$ are the symmetrized redshift space kernels for the galaxy density field.

The biased and RSD-corrected kernels are (see e.g.~\cite{Ivanov:2019pdj})
\begin{align}
    Z_1({\bf k}) =&\left(1+ \mu^2 \beta(k)\right) b_1\,,
    \label{z1}\\
    Z_{2}(\mathbf{k}_1,\mathbf{k}_2)  = \,&b_1 \bigg\{ F_{2}(\mathbf{k}_{1},\mathbf{k}_{2})+\beta(k)\mu^{2}G_{2}(\mathbf{k}_{1},\mathbf{k}_{2})\nonumber \\
    & +b_1 \frac{\beta(k) \mu k}{2}\left[\frac{\mu_{1}}{k_{1}}\big(1+\beta(k_{2})\mu_{2}^{2}\big)+\frac{\mu_{2}}{k_{2}}\big(1+\beta(k_{1})\mu_{1}^{2}\big)\right]\bigg\}+\frac{b_{2}}{2}+b_{{\cal G}_2}S_{1}(\mathbf{k}_{1},\mathbf{k}_{2})\,,
\end{align}
(already symmetrized) and
\begin{align}
    Z_{3}(\mathbf{k}_{1},\mathbf{k}_{2},\mathbf{k}_{3})
    \!= \, & b_1\bigg\{ F_{3}(\mathbf{k}_{1},\mathbf{k}_{2},\mathbf{k}_{3})+\beta(k) \mu^{2}G_{3}(\mathbf{k}_{1},\mathbf{k}_{2},\mathbf{k}_{3})\nonumber\\
    &\;+\! b_1 \beta(k) \mu k \frac{\mu_3}{k_3}\Big[ F_{2}(\mathbf{k}_{1},\mathbf{k}_{2})+\beta(k_{12})\mu_{12}^{2}G_{2}(\mathbf{k}_{1},\mathbf{k}_{2})\Big]\nonumber \\
    &\;+\! b_1 \beta(k) \mu k \big(1+\beta(k_1) \mu_{1}^{2}\big)\frac{\mu_{23}}{k_{23}}G_{2}(\mathbf{k}_{2},\mathbf{k}_{3}) \!+\! b_1^2 \frac{\big[\beta(k) \mu k\big]^{2}}{2}\big(1+\beta(k_1) \mu_{1}^{2}\big)\frac{\mu_{2}}{k_{2}}\frac{\mu_{3}}{k_{3}}\bigg\}\nonumber \\
    & \!+2b_{{\cal G}_2}S_{1}(\mathbf{k}_{1},\mathbf{k}_{2}+\mathbf{k}_{3})F_{2}(\mathbf{k}_{2},\mathbf{k}_{3})+b_1 b_{{\cal G}_2}\beta(k)\mu k\frac{\mu_{1}}{k_{1}}S_{1}(\mathbf{k}_{2},\mathbf{k}_{3})\nonumber \\
    & \!+2b_{\Gamma_3}S_{1}(\mathbf{k}_{1},\mathbf{k}_{2}+\mathbf{k}_{3})(F_{2}(\mathbf{k}_{2},\mathbf{k}_{3})-G_{2}\big(\mathbf{k}_{2},\mathbf{k}_{3})\big)\,,
 \label{z3}
\end{align}
(to be symmetrized), where ${\bf k} ={\bf k}_1+{\bf k}_2$ in $Z_2$ and   ${\bf k} ={\bf k}_1+{\bf k}_2+{\bf k}_3 $ in $Z_3$, $\mu_i\equiv \mu({\bf k}_i)$ ($i=1,\dots, 3$) are the angle cosine between ${\bf k}$ and the line of sight, $k_{12}=|{\bf k}_1+{\bf k}_2|$, $\mu_{12}\equiv \mu({\bf k}_{12})$, and so on.  $F_{2,3}$ and $G_{2,3}$ are the density and velocity kernels at second and third order, respectively, in standard perturbation theory.

The biasing scheme that includes all possible operators up to third order is \cite{2020PhRvD.102f3533C}:
\begin{equation}
    \delta_g=b_1\delta+\frac{b_2}{2}\delta^2+b_{\mathcal{G}_2}\mathcal{G}_2+b_{\Gamma_3}\Gamma_3,
\end{equation}
where
\begin{equation}
    \mathcal{G}_2(\Phi_g)\equiv(\partial_i\partial_j\Phi_g)(\partial^i\partial^j\Phi_g)-(\partial_i^2\Phi_g)^2
\end{equation}
and
\begin{equation}
    \Gamma_3\equiv\mathcal{G}_2(\Phi_g)-\mathcal{G}_2(\Phi_v)\,,
\end{equation}
with the gravitational potential $\Phi_g$ and the velocity potential $\Phi_v$ were used.
We also used
\begin{equation}
    S_{1}(\mathbf{k}_{1},\mathbf{k}_{2}) \equiv \Bigg[\frac{(\mathbf{k}_1\cdot\mathbf{k}_2)^2}{k_1^2k_2^2}-1\Bigg]
\end{equation}
and $\beta (k) \equiv f(k)/b_1$ with the growth rate \cite{2022MNRAS.512.2841Q}
\begin{equation}
    f\equiv\frac{\dd \log\delta}{\dd \log a}=-\frac{\dd\log D_{+}(z)}{ \dd\log(1+z)}\approx\Omega_m(z)^\gamma \,.
\end{equation}

The kernels  we adopted in this work have been derived for an Einstein-deSitter cosmology.
They include three free $z$-dependent functions, denoted $b_1$, $b_2$, $b_{\mathcal{G}_2}$. The bias parameter $b_{\Gamma_3}$ is set to zero. The restriction to Einstein-deSitter is of course a break of the model-independence, although it is well-known that this approximation works very well also for $\Lambda$CDM and other cosmologies that do not depart too much from it. In our FreePower method the kernels have been implemented in a much more general way, under general conditions like equivalence principle and extended Galilean invariance, and can be considered essentially model-independent.

The counterterms are added to take into account the rotational part of the flow and improve convergence. They can be parametrized as follows \cite{2020PhRvD.102f3533C}:
\begin{equation}
    P_{\rm ctr}(k,\mu,z)=-2\big(c_0(z) - (f/3)c_2(z)  \big)k^2P_{\rm L}(z,k) - 2 c_2(z) f(z) \mu^2k^2P_{\rm L}(z,k).
\end{equation}
The term in $c_0$ only enters the monopole, while the $c_2$ terms only enter the quadrupole. A next-to-next-leading order term, of the form $ -\tilde{c}b_{1}^{2}f^{4}\mu^{4}k^{4}(1+\beta\mu^{2})^{2}P(k)$ (see ISZ20) should also be included to correct for the Fingers-of-God effect at high $k$. We neglected it here for simplicity, and also because it was not documented in the version of {\tt pybird} that we used when we started this analysis. We plan to include it in future analyses.

The stochastic contribution to the power spectrum is modeled as a constant shot noise or Poisson noise. The number density in BOSS is $\Bar{n}\approx 3\times 10^{-4} (h/{\rm Mpc})^3$ \cite{Ivanov:2019pdj} which leads to a Poisson noise of $1/\Bar{n}$. However, deviations from this value are possible through exclusion effects \cite{2013PhRvD..88h3507B}, and therefore we include the variable shot-noise parameter defined in Eq.~\eqref{eq:shot}. The multipoles can be computed via
\begin{equation}
    P_{{\rm gg},l}(z,k)\equiv\frac{2l+1}{2}\int_{-1}^1{\rm d}\mu \,P_{\rm gg}(k,\mu,z)\mathcal{P}_l(\mu),
\end{equation}
where $\mathcal{P}_l(\mu)$ are the Legendre polynomials.

When the galaxy spectra are estimated from data, one has to assume a reference cosmology (fiducial cosmology). The fiducial cosmology, a flat $\Lambda$CDM cosmology with $\Omega_{m0}=0.31$, is used to convert observables (RA, DEC, $z$) into Cartesian coordinates. Then, quantities like $k$, $\mu$, and volumes are computed assuming the reference cosmological model. The reference cosmology is different from the true cosmology and this difference leads to a geometrical distortion, the Alcock-Paczyński effect. The observable galaxy power spectrum is \cite{Ivanov:2019pdj}
\begin{equation}
    P_{\rm obs}(k_{\rm obs},\mu_{\rm obs}) = P_{gg}\big(k_{\rm true}[k_{\rm obs},\mu_{\rm obs}],\mu_{\rm true}[k_{\rm obs},\mu_{\rm obs}]\big) \times \frac{L_{\rm A,fid}^2 E_{\rm true}}{E_{\rm A, true}^2 H_{\rm fid}}.
\end{equation}
Here $k_{\rm obs}$ and $\mu_{\rm obs}$ are the quantities  obtained under a given cosmological reference model. This is also the case for $L_{\rm A, fid}$ and $E_{\rm fid}$. To convert between the true quantities and the quantities derived in the reference model, it is necessary to know $E_{\rm true}$ and $L_{\rm A, true}$:
\begin{equation}
    k_{\rm true}^2=k_{\rm obs}^2\Bigg[\bigg(\frac{E_{\rm true}}{E_{\rm fid}}\bigg)^2 \mu_{\rm obs}^2 + \bigg(\frac{L_{\rm A, fid}}{L_{\rm A, true}}\bigg)^2 (1-\mu_{\rm obs}^2)\Bigg],
\end{equation}
\begin{equation}
    \mu_{\rm true}^2=\bigg(\frac{E_{\rm true}}{E_{\rm fid}}\bigg)^2 \mu_{\rm obs}^2\Bigg[\bigg(\frac{E_{\rm true}}{E_{\rm fid}}\bigg)^2 \mu_{\rm obs}^2 + \bigg(\frac{L_{\rm A, fid}}{L_{\rm A, true}}\bigg)^2 (1-\mu_{\rm obs}^2)\Bigg]^{-1}.
\end{equation}
When the Alcock-Paczyński effect is considered as well, the expression becomes
\begin{equation}
    P_{{\rm gg},l}(k)\equiv\frac{2l+1}{2}\int_{-1}^1{\rm d}\mu_{\rm obs} P_{\rm obs}(k_{\rm obs},\mu_{\rm obs})\mathcal{P}_l(\mu_{\rm obs}).
\end{equation}

\section{Interpolation schemes}\label{app:interpolation}

\begin{figure}
    \centering
    \includegraphics[width=0.65\columnwidth]{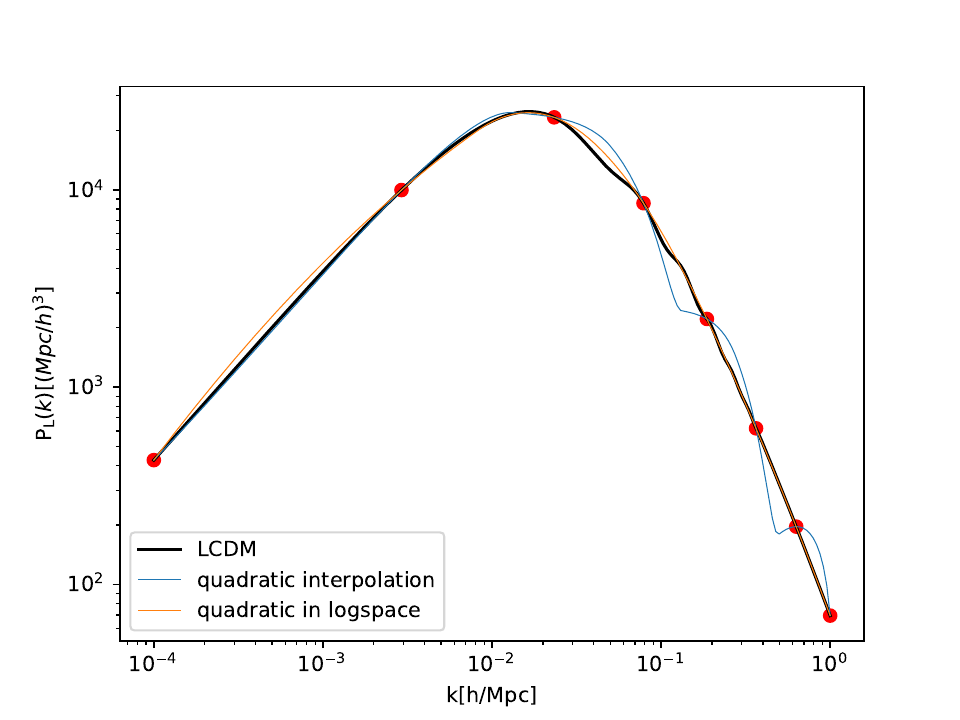}
    \caption{Visual comparison of two different quadratic interpolation schemes assuming a $\Lambda$CDM fiducial power spectrum.}
    \label{fig:interpolation-test}
\end{figure}
\begin{table}
    \setlength\tabcolsep{3.3pt}
    \centering
    \begin{tabular}{lccc}
    \hline
    interpolation scheme & $\sigma_8$ ($8$ bins) & $\sigma_8$ ($16$ bins) \\
    \hline
    fiducial value & \multicolumn{2}{c}{$0.823$} \\
    linear & $0.909$ & $0.835$ \\
    quadratic & $0.828$ & $0.818$ \\
    quadratic in logspace & $0.831$ & $0.819$\\
    cubic & $0.839$ & $0.819$ \\
    \hline
    \end{tabular}
    \caption{Comparison of values for $\sigma_8$ at $z=0$ obtained by using various interpolation schemes and $k$-bin numbers.
    }
    \label{tab:sig8_interpolation_schemes}
\end{table}

To investigate the effect of the interpolation scheme and the number of $k$-bins on the obtained values for $\sigma_8$, we use a fiducial linear matter power spectrum obtained assuming the $\Lambda$CDM model. For our low number of $k$-bins, this choice can have a significant impact on $\sigma_8$. The highest discrepancy, of around 10\%, is seen when using a simple linear interpolation, as shown in Table~\ref{tab:sig8_interpolation_schemes}. A simple quadratic interpolation reduces the discrepancy to around $0.6\%$.

However, the final form of $P(k)$ becomes spuriously wiggled, as can be seen in Figure~\ref{fig:interpolation-test}. We therefore adopted an interpolation which is instead quadratic in logspace. This is a good compromise, as the resulting $P(k)$ is similar to the fiducial one, while $\sigma_8$ is still recovered with a small $1\%$ bias. 

This dependency on the interpolation choice is naturally smaller when increasing the number of $k$-bins. We find that using 16 bins leads to a $1.5\%$ or better agreement with the fiducial $\sigma_8$ for all interpolation choices, at the cost of a higher dimensionality for the MCMC chains. This can be further explored in the future when better datasets allow for a more precise estimate of $\sigma_8$ with the FreePower method.

\section{Effect of hexadecapole in the matrix mixing}\label{app:hexadecapole_matrix_mixing}

\begin{figure}
    \centering
    \includegraphics[width=0.65\columnwidth]{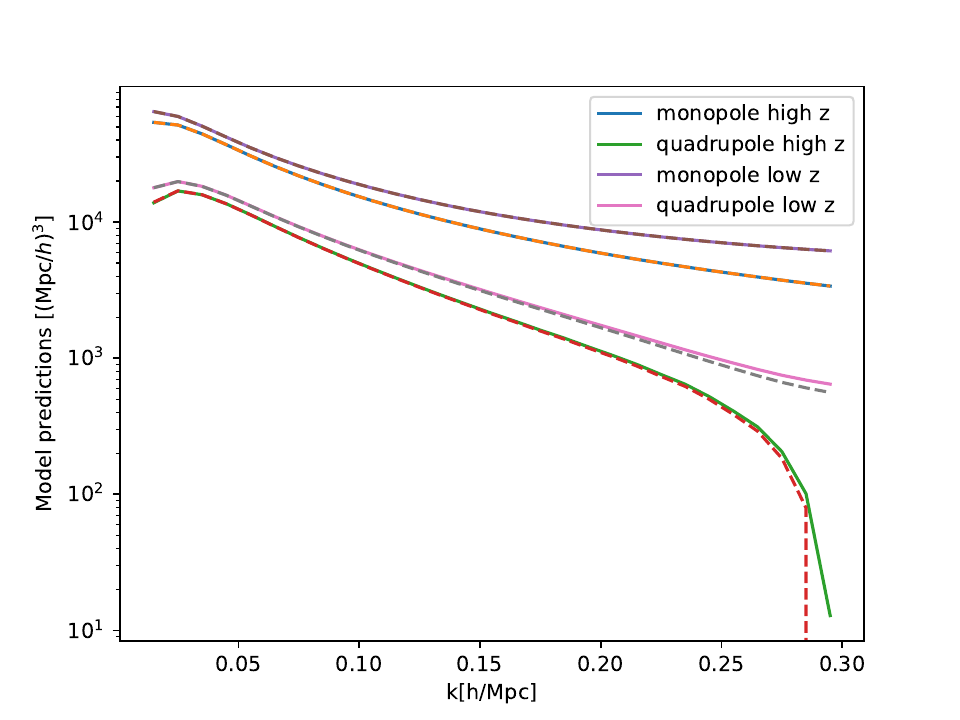}
    \caption{Model predictions for monopole and quadrupole and again when the hexadecapole is not set to zero. The differences arise due to the mixing matrix multiplication.}
    \label{fig:mixing_matrix_multiplication}
\end{figure}

In our analysis we set the theoretical hexadecapole to zero. However, due to the mixing matrix which multiplies the monopole and quadrupole data could in principle be affected by this choice. Figure~\ref{fig:mixing_matrix_multiplication} shows the model predictions for monopole and quadrupole when the hexadecapole is either set to zero or not. The model values are evaluated at our our inferred parameter values (see Table~\ref{tab:main-results}). No effect in the monopole can be seen, while only a small correction is seen for the quadrupole, and then only for $k > 0.2$ Mpc/$h$. We conclude that this choice for the hexadecapole has minimal impact on our present analysis.

\bibliographystyle{JHEP}
\bibliography{literature}

\end{document}